\documentclass[12pt,notitlepage,a4paper]{article}
\pdfoutput=1
\usepackage{color,graphicx}
\usepackage{cite}
\usepackage{amssymb}
\usepackage{delarray,amsmath,bbm}
\usepackage[latin1]{inputenc}
\usepackage[american]{babel}
\usepackage{cite}
\usepackage{float}

\newcommand{\be}{\begin{equation}}
\newcommand{\ee}{\end{equation}}
\newcommand{\beq}{\begin{equation}}
\newcommand{\eeq}{\end{equation}}
\newcommand{\beqa}{\begin{eqnarray}}
\newcommand{\eeqa}{\end{eqnarray}}

\newcommand{\bear}{\begin{eqnarray}}
\newcommand{\eear}{\end{eqnarray}}

\newcommand{\pd}{\partial}
\numberwithin{equation}{section}

\newfont{\namefont}{cmr10}
\newfont{\addfont}{cmti7 scaled 1440}
\newfont{\boldmathfont}{cmbx10}
\newfont{\headfontb}{cmbx10 scaled 1728}

\begin{document}
\baselineskip=15.5pt
\pagestyle{plain}
\setcounter{page}{1}

\begin{center}
%\%\rightline{US-FT-3/07}
\vspace{0.1in}

\renewcommand{\thefootnote}{\fnsymbol{footnote}}

\begin{center}
\Large \bf  Stability of Charged Global  AdS$_4$ Spacetimes
\end{center}
\vskip 0.1truein
\begin{center}
\bf{Ra\'ul Arias,${}^1$\footnote{rarias@cab.cnea.gov.ar}
Javier Mas,${}^2$\footnote{javier.mas@usc.es}
Alexandre Serantes${}^2$\footnote{alexandre.serantes@usc.es}}\\
\end{center}
\vspace{0.5mm}

\begin{center}\it{
${}^1$Centro At\'omico Bariloche\\CONICET\\
8400-S.C. de Bariloche, R\'{\i}o Negro, Argentina}
\end{center}

%\begin{center}
\begin{center}\it{
${}^2$Departamento de  F\'\i sica de Part\'\i  culas \\
Universidade de Santiago de Compostela \\
and \\
Instituto Galego de F\'\i sica de Altas Enerx\'\i as (IGFAE)\\
E-15782 Santiago de Compostela, Spain}
\end{center}

\setcounter{footnote}{0}
\renewcommand{\thefootnote}{\arabic{footnote}}

\vspace{0.4in}

\begin{abstract}
\noindent
We study linear and nonlinear stability of asymptotically  AdS$_4$ solutions in Einstein-Maxwell-scalar theory. After summarizing the set of static solutions
 we first examine thermodynamical stability in the  grand canonical ensemble and the phase transitions that occur among them.
In the second part of the paper we focus on nonlinear stability in the microcanonical ensemble by evolving radial perturbations numerically.  We find hints of an instability corner for vanishingly small perturbations of the same kind as the ones present in the uncharged case. Collapses are avoided, instead, if the charge and mass of the perturbations come to close the line of solitons. Finally we examine the soliton solutions.  The linear spectrum of normal modes is not resonant and  instability turns on at extrema of the mass curve.
Linear stability extends to nonlinear stability up to some threshold for the amplitude of the perturbation.
Beyond that, the soliton is destroyed and collapses  to a hairy black hole.
The relative width of this stability band scales down with the charge Q, and does not survive the blow up limit to a planar geometry.

\end{abstract}

\smallskip
\end{center}

\newpage

\tableofcontents

\break

\section{Introduction}

Stability of Anti de Sitter vacua has received much attention in the last years since the numerical experiment  in \cite{Bizon:2011gg} . There it was found that
nonlinear evolution of some family of arbitrarily small scalar field perturbations inevitably end up in the collapse and formation of a black hole.  This is the so called instability corner \cite{Dimitrakopoulos:2014ada}. Perturbatively, the problem was also examined  \cite{Dias:2011at}\cite{Dias:2012tq} in the context of purely gravitational perturbations,  and the importance of two ingredients was signalled: the presence of a fully resonant spectrum for the linearized perturbations, and the existence of  periodic solutions that  acted as centers of some stability islands in the space of initial conditions. Back to the scalar field case, this suggestion received further backup from other contributions \cite{Buchel:2013uba}\cite{Maliborski:2014rma}. In this case, the periodic solutions were named oscillons, and indeed, the two previous observations became consistent in that the spectrum of linearized perturbations around an oscillon turns out not to be fully resonant. After some years of analytic and numerical work, it has become clear that there is a wealth of situations that one can encounter. One may choose to change either the dynamics (the action) or the kinematics (the boundary conditions). Generically, when departing from the easiest case of perturbations on pure AdS, the resonant property is lost \cite{Maliborski:2012gx}\cite{Deppe:2014oua}. An odd case is that of AdS$_3$, where the mass gap (of conical singularities) kills the instability corner despite the fact that the spectrum is fully resonant \cite{Bizon:2013xha}\cite{daSilva:2014zva}.

In this paper we study both linear and nonlinear stability of asymptotically AdS vacua in Einstein-Maxwell theory  interacting with a charged massless scalar.  There's little we can add   to motivate focussing on this theory.
The space of static solutions is rich and involves regular solitons, hairy and Reissner-Nordstr\"om (RN) black holes. The first two are interpreted in the context of AdS/CFT as holographic duals of quantum states with  spontaneously broken global $U(1)$ symmetry.  We will be working in AdS$_4$ in global coordinates. Our findings build up in parallel with  the achievements of  \cite{Basu:2010uz}  \cite{Dias:2011tj} where the landscape of static solutions was unraveled for the case of AdS$_5$. The thermodynamical (linear) stability issue concerning the competition of these three solutions depends upon the ensemble that is considered. We will construct the grand potential to find the correct vacua in the grand canonical ensemble.

For the nonlinear stability we must carry out  numerical evolution analysis, searching for endpoints of the evolution in one of the above possible static forms. Performing these simulations with Dirichlet boundary conditions at the boundary, is tantamount to studying the thermalization of the dual quantum system in the microcanonical ensemble.

The Lagrangian that governs the dynamics is given by\footnote{We find it advantageous to keep this somewhat old fashioned normalization, since it allows to make connection with different conventions in the literature. For example, the one used \cite{Basu:2010uz} \cite{Dias:2011tj}  is recovered by setting $\kappa^2=1$.
In this paper we will set instead $\kappa^2=(d-1)/2$.
 }
\be
S= \int d^{d+1}x \sqrt{-g}\left[ \frac{1}{2\kappa^2} \left({\cal R} + \frac{d (d-1)}{l^2} \right)   -\frac{1}{4} F_{\mu\nu} F^{\mu\nu}  -   | D_\mu\phi |^2  \right]
\ee
for an Anti de Sitter space with curvature radius $l$.  Here $D_\mu \phi =(\pd_\mu - i e A_\mu)\phi$ and there is a $U(1)$ gauge symmetry. The coupling $e$ is a free parameter and we can measure all lengths in units of $l$, hence setting $l=1$. Analytic expressions for generic $d$ are provided in the appendices, but the numerical analysis will examine only the $d=3$ case.

RN and hairy black holes are dual to normal and superfluid phases of the dual field theory respectively. In this last case, it is the scalar  field that condenses, breaking spontaneously the $U(1)$ global symmetry.  The dual state is called an $s$-wave superfluid \cite{Hartnoll:2008kx}.
  A similar  setup was considered in \cite{Gubser:2008px}, albeit in the probe limit, while our analysis takes full account of the backreaction of the fields on the geometry. In this context, it is known that black holes come typically in pairs, one of which is small and thermodynamically unstable. Moreover, there are solitonic solutions which are completely regular in the bulk. Whenever these are the ground states, they should correspond in the dual field theory to zero entropy condensates. Finally, we also find extremal black holes, both among RN and hairy solutions.

The plan of the paper is the following. Section 2 contains a brief summary of the set of solutions that can be obtained as a function of the electromagnetic coupling $e$. It is written in the same style as in \cite{Basu:2010uz} and \cite{Dias:2011tj} for AdS$_5$, highlighting differences and similarities. In section 3 we address the issue of the connection of such solutions with superfluid states. This requires identifying the thermodynamically dominant phase in the grand canonical ensemble. We compute the renormalized grand potential and find out both first and second order phase transitions. nonlinear stability is the topic of concern in section 4. It requires numerically evolving arbitrarily small initial perturbations over AdS$_4$. The central question in this section is whether similar conclusions as those  obtained in \cite{Bizon:2011gg} can be extrapolated to the present situation. The answer points in the affirmative, namely, the vacuum exhibits a corner of nonlinear instability even at finite charge. On the other hand, the islands of stability get amplified, probably as a consequence of the electrostatic repulsion.
This matches with expectations, since the AdS$_4$ linear scalar perturbations are still fully resonant in this theory. In section 5 we study the stability of solitonic ground states. In a band above them,  oscillating solutions never decay. We found the upper bound of this protected from collapse region and examine the way it scales upon blowing up into a black brane geometry. Our findings indicate that, in such limit,  this region does not survive. We also determine the scalar field normal mode spectrum above a solitonic background, finding that it has a dispersive character. In the light of findings in \cite{Dias:2012tq} this implies that the weakly turbulent instability present for the AdS$_4$ vacuum is absent in these geometries.

\section{Static solutions}

In this section we describe the space of static solutions as a function of the coupling $e$. We shall start by providing some background material, and relegate to the appendices all the cumbersome expressions.

\subsection{Construction}

The  ansatz for the metric of AdS$_{d+1}$ in global coordinates  is the following
\be
ds^2 = \frac{l^2}{\cos^2 x}\left( - f(t,x) e^{-2\delta(t,x)} dt^2+ f(t,x)^{-1} dx^2 + \sin^2 x \, d \Omega_{d-1}^2\right),  \label{line1}
\ee
where $x\in [0,\pi/2]$. The standard Schwarzschild radial coordinate $r$ is related to $x$ as $r = \tan(x)$. Concerning the gauge field,  isotropy is maintained by selecting
$
A = A_t(t,x) dt  .
$
This ansatz leaves as residual gauge symmetry
\be
\phi \to e^{i\Lambda(t)}\phi, ~~~~ A_t \to A_t + \frac{1}{e} \pd_t \Lambda(t) \, ,  \label{resgauge}
\ee
under which the equations of motion must be covariant. This motivates defining the following  $U(1)$ covariant fields\footnote{Given some function $g = g(t,x)$, we define $\dot{g}\equiv\partial_t g$ and $g'\equiv\partial_x g$.}
\beqa
&&\Phi(t,x) \equiv \phi'(t,x), \\
&&\Pi(t,x) \equiv \frac{e^\delta}{f} D_t\phi \, .
\eeqa
The ansatz \eqref{line1} breaks general covariance leaving a residual time-reparametrization symmetry
\be
\delta \to \delta - \chi \,~~~~~~~A_t\rightarrow e^{\chi} A_t,~~~~~t\rightarrow e^{-\chi} t\, . \label{scaling}
\ee
This can be employed to either set $\delta_{b}$ or $\delta_{o}$ to zero, fixing the time coordinate $t$ to be the proper time at either $x_b=\pi/2$ or $x_o$, respectively.

Equations of motion for the time dependent ansatz can be found in appendix A, eqs. \eqref{eqmovPhi}-\eqref{Einstein2} and \eqref{intat} . Equations for static solutions follow  after setting $\Pi = - i e^{\delta}A_t \phi/f$, $\dot\Pi=0$ and $C=A_t'$. We provide them here for  completeness
 \begin{eqnarray}
&&\phi''(x) + \left(\left(d-1\right)\csc(x)\sec(x) + \frac{f'(x)}{f(x)} - \delta'(x)\right) \phi'(x) + e^2\frac{ e^{2 \delta(x)}A(x)^2 \phi(x)}{f(x)^2} = 0, \nonumber \\
&&A''(x) + \left(\left(d-1 \right)\cot(x) + \left(d-3\right)\tan(x) + \delta'(x)\right) A'(x) -  e^2\frac{2 \sec(x)^2 A(x) \phi(x)^2}{f(x)} = 0, \nonumber  \\
&&f'(x) - \frac{d-2+2\sin(x)^2}{\sin(x)\cos(x)}(1-f(x)) - f(x) \delta'(x) + \frac{1}{2}\cos(x)^3 \sin(x) e^{2\delta(x)}A'(x)^2 = 0,  \nonumber \\
&&\delta'(x) + \cos(x)\sin(x)\left(\phi'(x)^2 + e^2\frac{ e^{2\delta(x)}A(x)^2 \phi(x)^2}{f(x)^2} \right) = 0. \label{static_eqs}
\end{eqnarray}
The scalar field $\phi$ can be taken  real by an appropriate gauge choice. Solving \eqref{static_eqs} requires, as usual, specifying boundary conditions at both ends of the radial domain. At the boundary, $x_b =\frac{\pi}{2}$, which is holographically related to the UV regime of the field theory, the following asymptotics follow directly from the equations of motion, setting $\rho=x-\frac{\pi}{2}$,
\beqa
\phi(x)&=&\phi_{b} + \ldots +\phi_{b,3}\rho^3+\ldots, \label{bexp1}\\
A_t(x)&=&\mu+Q \rho+\ldots,\\
f(x)&=&1 + \ldots + M\rho^3+\ldots,\\
\delta(x)&=& \delta_{b} + \ldots +  \delta_{b,4} \rho^4+\ldots \, . \label{bexp4}
\eeqa
 The holographic dictionary identifies the leading term of the scalar field asymptotic series expansion, $\phi_b$, with the source of a marginal field theory operator, ${\cal O}$, and the subleading term $\phi_{b,3}$  with its vacuum expectation value,  $\langle{\cal O}\rangle$. Likewise, from the asymptotic expansion of the gauge field we can read off both the chemical potential, $\mu$, and charge density, $Q \propto \langle J^t\rangle$, of the dual field theory state.

We refer to the infrared end of the radial coordinate as $x_{o}$, with either $x_o=0$ for solitonic solutions, or $x_o=x_h > 0$ for the horizon of a black hole. The infrared series expansions read
\beqa
\phi(x)&=&\phi_o+\phi_{o,1}(x-x_{o})+\ldots,   \label{serphi0} \\
A_t(x)&=&A_o   + A_{o,1}(x-x_o)+ \ldots,  \label{serA0} \\
\delta(x)&=&\delta_o + \delta_{o,1}(x-x_o) +\ldots,  \label{serdelta0} \\
f(x)&=&f_o + f_{o,1}(x-x_o)+\ldots.  \label{serf0}
\eeqa
where it should be noted that regularity of solutions at the origin demands $f_o = 1$ and forces every odd term of \eqref{serphi0}-\eqref{serf0} to be zero. In the black hole case, the existence of a horizon and regularity of the gauge field one-form implies
\be
f_h =  A_h = 0\, . \label{condhor}
\ee

Each static solution of the equations of motion is completely characterized by its infrared series expansion which, in turn, is totally fixed in terms of a finite number of parameters. In the absence of a horizon, these parameters can be  the values of the scalar and gauge fields at $x_o=0$, $(\phi_0, A_0)$. A soliton geometry is dual to a field theory state with spontaneously, and not explicitly, broken symmetry. This further demands that the source $\phi_b$ vanishes, which provides a nonlinear relation between $(\phi_0, A_0)$ that determines completely $A_0$ in terms of $\phi_0$. A family of soliton solutions is therefore uniparametric. The same reasoning goes through to the black hole case. Now, each black hole solution would be totally determined by the triplet $(x_h, \phi_h, A_{h,1})$ and, again, the condition $\phi_b = 0$ would link $\phi_h$ and $A_{h,1}$ at the given $x_h$. In this way, a family of black hole solutions is bi-parametric.

%To end this subsection, let us note that the system \eqref{static_eqs} enjoys one residual gauge symmetry. This is a residual time reparametrisation invariance
%\be
%\delta \to \delta - \chi \,~~~~~~~A_t\rightarrow e^{\chi} A_t,~~~~~t\rightarrow e^{-\chi} t\label{scaling}
%\ee
%that can be employed to set either $\delta_{b}$ or $\delta_{o}$ to zero, fixing the time coordinate %$t$ to be the proper time at either $x_b=\pi/2$ or $x_o$, respectively. Moreover there is a residual gauge symmetry of the whole system that comes from \eqref{resgauge}, with $\Lambda(t) = \lambda t$. This transformation shifts $A_b$ and $A_o$ by a constant, in such a way that only the difference $A_b - A_o$ is gauge invariant.\footnote{This symmetry is absent in a black hole background, since it would change the holonomy of the gauge connection around the Euclidean time cycle.}

The backreaction of the gauge field on the geometry is controlled by the coupling $e$ and the probe limit corresponds to $e\rightarrow\infty$.

\begin{figure}[h]
\centering
\includegraphics[width=6cm]{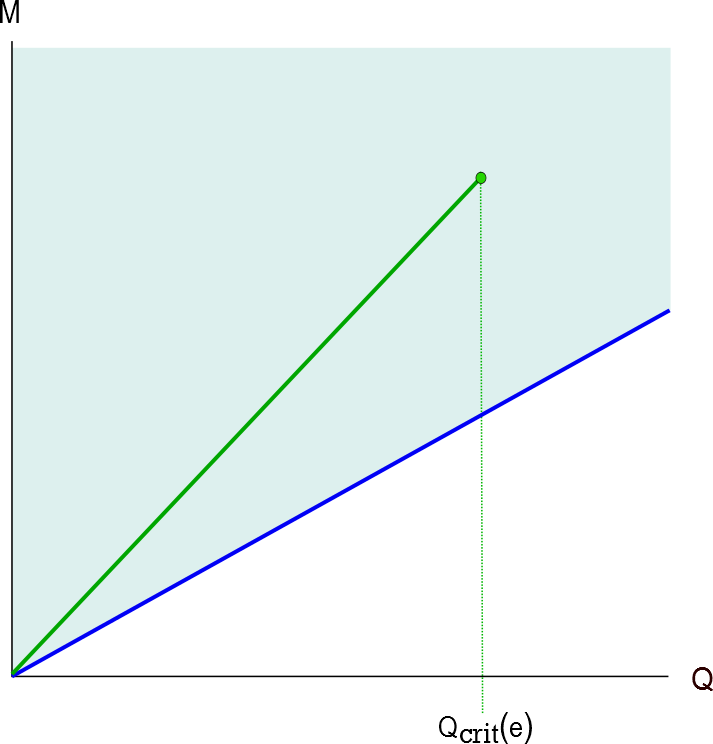}~~~~~~~~~~~~~~~~\includegraphics[width=6cm]{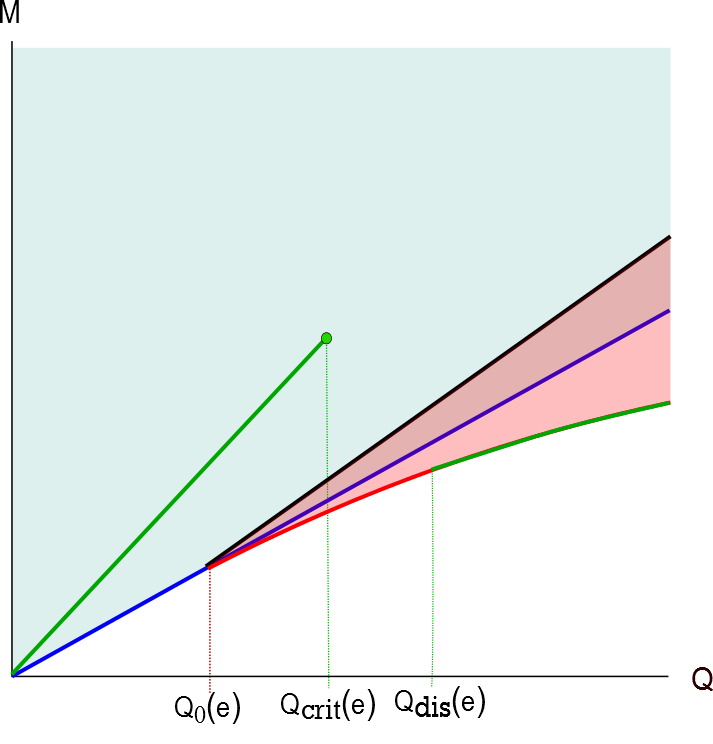}
\caption{{\small Left plot: phase diagram for $e<e_t$. In this range, there is a single soliton branch that ends up in a limiting value for the charge. The green shaded region are regular RN black holes, and they become extremal at the bottom blue line.  Right plot: in this intermediate regime, $e_t<e<e_{sr}$, hairy solutions appear in a band about extremality. The red line stands for extremal hairy black holes, below which there are no solutions. Here we have a critical limiting charge for the vacuum connected solitons (upper green curve) and an unbounded vacuum disconnected soliton line (lower green line).}}
\label{menor}
\end{figure}

\subsection{Summary of phases}

Following closely the  discussion contained in \cite{Basu:2010uz} \cite{Dias:2011tj}, we  parameterise the space of solutions at a given coupling, $e$,  by the charge $Q$ and mass $M$. Three  different regimes appear separated by two threshold values,  $e_t$ and $e_{sr}$,   that signal  the appearance of two distinct instabilities. The lower one, $e_t$, marks the threshold for  a near-horizon tachyonic instability that affects RN black holes. It
 is triggered by the fact that the gauge field of an extremal RN black hole makes a negative contribution to the effective scalar field mass, lowering it below the Breitenlohner-Freedman bound of the AdS$_2$ factor of the near-horizon geometry. It is of the same kind as the one found in the context of holographic superconductors \cite{Hartnoll:2008kx} \cite{Gubser:2008px}. As discussed in appendix B, it is only relevant for large enough black holes, a fact that explains why it was found in the context of planar geometries. On the other hand, for small black holes, there is a superradiant instability at work beyond $e_{sr}$ \cite{Basu:2010uz} \cite{Dias:2011tj}. In the following table we show the values of these thresholds for AdS$_{d+1}$ with $d=3,4$ for $\kappa^2 = 1$
$$
\begin{array}{|c|c|c|}\hline ~ &\rule{0mm}{4mm} e^2_{t}  & e^2_{sr} \\ \hline  {\rm AdS}_4 \rule{0mm}{4mm}&\rule{2mm}{0mm} 3/2 \rule{2mm}{0mm} &\rule{2mm}{0mm} 9/2 \rule{2mm}{0mm}  \\ \hline {\rm  AdS}_5 \rule{0mm}{4mm}& 3 & 32/3 \\ \hline \end{array}
$$
In this table, $e_t$ stands for the tachyonic instability threshold, and $e_{sr}$ for the superradiant one.\footnote{General expressions for this thresholds valid at any $d$ can be found in the appendix B.} Within each interval of possible values for the coupling we have encountered a similar phenomenology.

In the remaining part of this section we summarize the phase diagram for the static solutions of the Einstein-Maxwell-scalar action for values of the coupling lying on each interval and discuss the soliton solutions in those regimes.

\begin{description}
\item{ $\bullet ~~e<  e_t$}

Here, the coupling is below the tachyonic instability threshold. The only static solutions present are RN black holes and solitons which exist for values of the charge smaller than $Q_{crit}(e)$, a sort of Chandrasekar limit on regular selfgravitating solutions. In figure \ref{menor} (left) the green curve corresponds to solitons and the blue curve to the extremal RN black holes. Regular RN black hole solutions exist in the blue shaded region above this curve.

\item{ $\bullet ~~e_t< e < e_{sr}$}

In this regime, in addition to RN black holes and regular solitons, there exist large and small hairy black holes in a band about extremality for $Q> Q_0(e)$ (red shaded region of figure \ref{menor}, right).  The red curve denotes  extremal hairy black hole solutions and the black line is the instability curve for the RN solution.
This function has the limiting behaviours $\lim_{e\to e_t} Q_0(e) = \infty$ and $\lim_{e\to e_{sr}} Q_0(e) = 0$. Solitons, in green on figure \ref{menor}, appear in two branches. One is  connected with the vacuum and has bounded charge  $Q<Q_{crit}(e)$. The other one has instead a lower bound $Q>Q_{dis}(e)$. Both bounds merge  $Q_{dis}(e) \rightarrow Q_{crit}(e)$ in the upper limit $e\rightarrow e_{sr}$. When $Q>Q_{dis}(e)$ the lowest mass hairy BH is no longer extremal and has non vanishing (possibly diverging) temperature.
\item{ $\bullet ~~e_{sr} < e$}

\begin{figure}[htbp]
\begin{center}
\includegraphics[width=7cm]{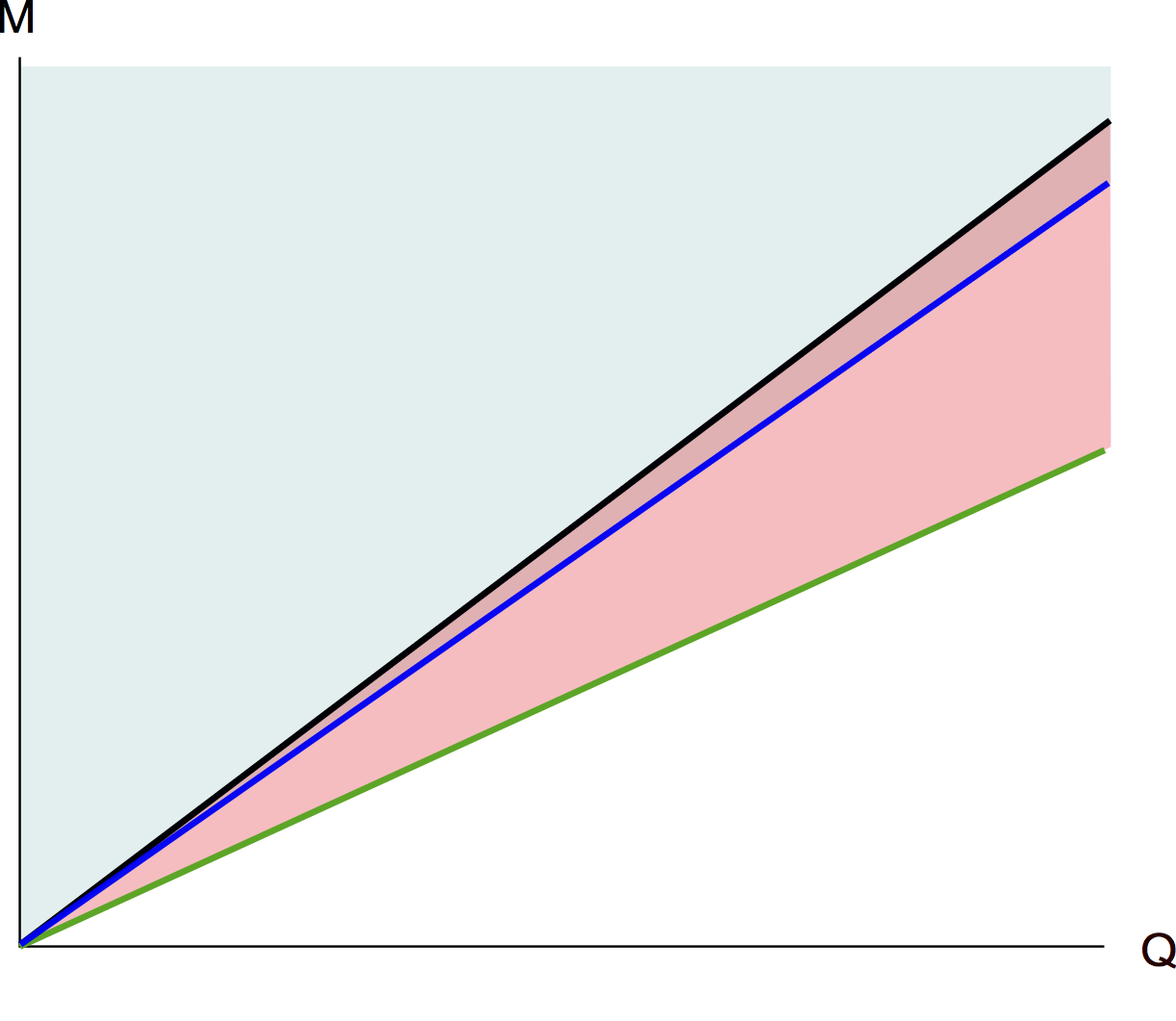}
\caption{\label{eupper} {\small For $e>e_{sr}$ we have two kind of instabilities that lead to the formation of hairy black holes, tachyonic and superradiant. Note that in this regime there exist solitonic solutions for any value of the charge.}}
\label{intermedio}
\end{center}
\end{figure}

Now hairy black holes exist for all values of $Q$ in a band about extremal RN. Concerning solitonic solutions, the two branches seen in figure \ref{menor} merge to form a single branch that extends unbounded for all values of $Q\geq 0$. To the best accuracy of our numerics, the soliton line is the limit of vanishingly small hairy black holes with diverging temperature. In this limit, the temperature becomes ill defined for the soliton.
The situation is very similar to the one in  AdS$_5$  \cite{Dias:2011tj}. However, in that case,  a second critical charge was found, beyond which hairy black holes appeared below the soliton line. We have not seen any trace of this behaviour in our numerical scan over the space of static solutions. However we cannot discard a fine structure of the form observed in AdS$_5$  beyond the reach of our numerical accuracy. For that, a perturbative analytical study along the lines used in \cite{Dias:2011tj} would be needed.

\end{description}

\begin{figure}[h!]
\begin{center}
\includegraphics[width=15.7cm]{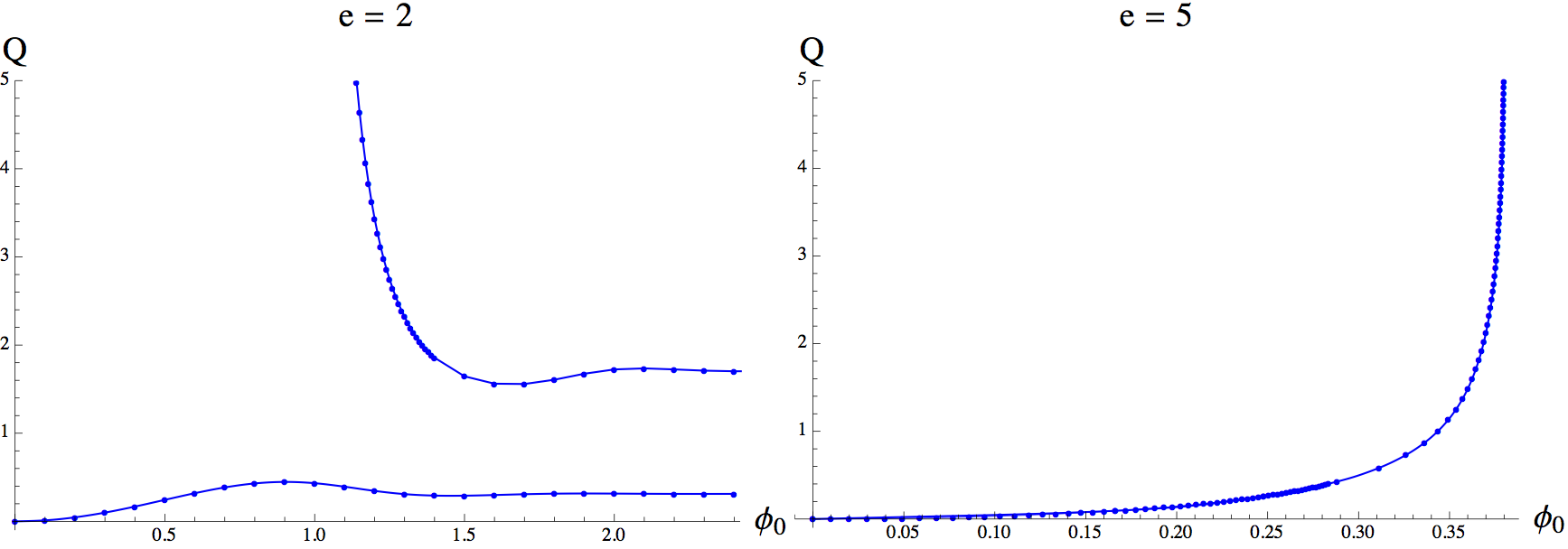}
\caption{\label{fig1} {\small Left: charge of vacuum connected and disconnected soliton branches against $\phi_0$ for $e = 2$. Right: soliton charge against $\phi_0$ for $e=5$.}}
\end{center}
\end{figure}

\subsubsection{Soliton branches in more detail}

\noindent Let us describe the soliton solutions in more detail. They are horizonless, fully backreacted, charged solutions, sourced by a normalizable scalar field profile $\phi_s(x)$, where equilibrium is attained by an exact compensation between gravitational an electric forces. Due to the normalizability condition, on the field theory side, a soliton represents a macroscopic Bose-Einstein condensate that spontaneously breaks the  $U(1)$  symmetry. They come in one-parameter families that can be parametrized by the central value $\phi_0 \equiv \phi_s(x = 0)$.
Depending on the value of the coupling $e$, soliton families display  different aspects\footnote{Our results agree essentially with the ones analyzed in \cite{Gentle:2011kv} for the case of  $m^2 = -2$.}
\begin{itemize}
\item For $e < e_t$, there exists a single soliton branch. It is continuously connected to the AdS$_4$ vacuum, in the sense that, for $\phi_0 \rightarrow 0$, it reduces to global AdS$_4$. Therefore, for $\phi_0 \ll 1$, this solution family admits a perturbative construction, and can be described as a nonlinearly dressed $\omega = 0$ AdS scalar normal mode. Besides this fact, the trademark  of this  branch is the existence of a critical value  $\phi_0 = \phi_{c,1}$ for which both the soliton mass $M(\phi_0)$ and charge $Q(\phi_0)$ attain a maximum. When $\phi_0 > \phi_{c,1}$, $M$ and $Q$ spiral around a limiting value that is reached for $\phi_0 \rightarrow \infty$.
\item For $e_{t} < e < e_{sr}$, there exist two different soliton branches. The first one, connected with the vacuum,  was already present in the $e < e_t$ case. The second one, disconnected from the  vacuum,   is not amenable to a perturbative construction. In this branch, solutions exist for $\phi_0$ larger than some critical value $\phi_{c,2a}$ for which the conserved charges $(M,Q)$ diverge. They decrease with $\phi_0> \phi_{c,2a}$ until they reach a minimum value at some $\phi_0=\phi_{c,2b} > \phi_{c,2a}$.
In parallel with the  first soliton branch, for $\phi_0 > \phi_{c,2b}$, $M$ and $Q$ show damped oscillations around a limiting value that is attained in the $\phi_0 \rightarrow \infty$ limit.
Representative plots of the behavior just described are provided in figure \ref{fig1} (left).
\item For $e \geq e_{sr}$, the two soliton branches described in the previous item fuse into a single solution family that is vacuum connected (see figure \ref{fig1}, right). Again, there exists a critical $\phi_{c,3}$ such that $M$ and $Q$ seem to diverge  in the limit $\phi_0 \rightarrow \phi_{c,3}$.
\end{itemize}

\begin{figure}[h!]
\begin{center}
\includegraphics[width=13cm]{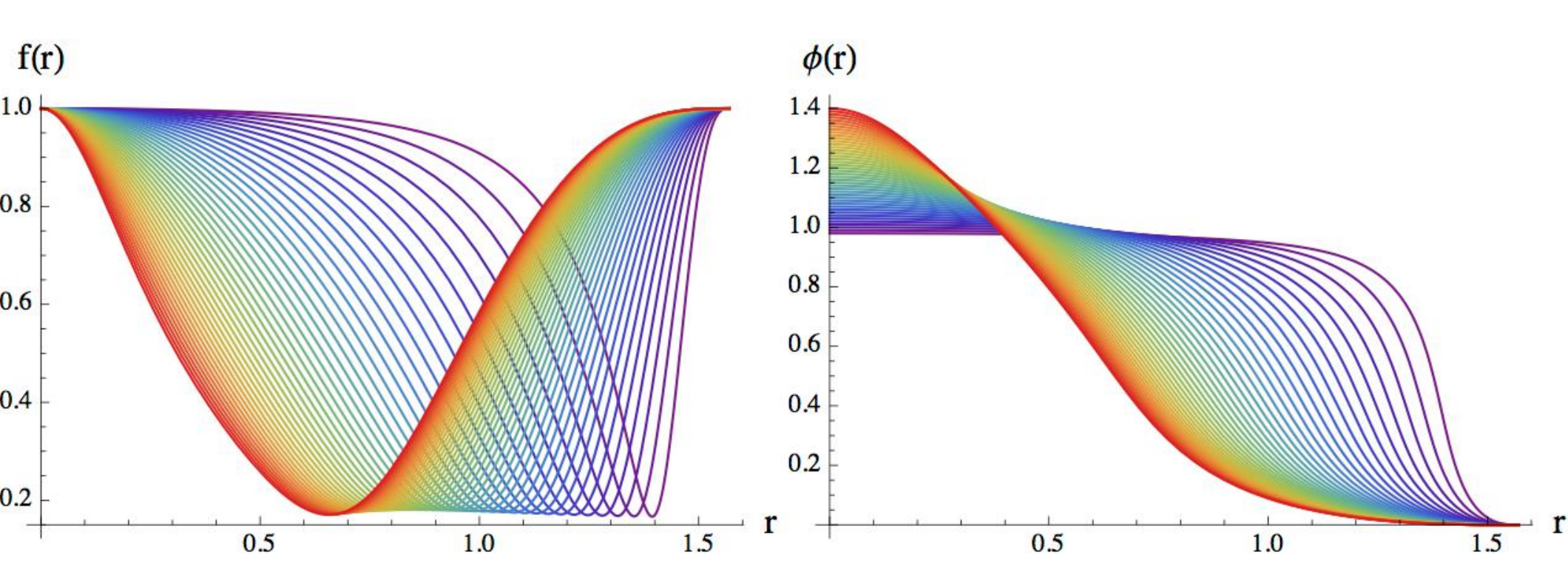}
\caption{\label{fig2} {\small Fields for the vacuum disconnected soliton branch at $e = 2$ from $\phi_0 = 1.4$ (red) to $\phi_0 = 0.98$ (magenta) in steps of $\delta\phi_0 = -0.01$.}}
\end{center}
\end{figure}

\subsubsection{The blow up limit}

\noindent In figure \ref{fig2} we plot $f_s$ and $\phi_s$ for representative solutions of the unbounded branch at $e = 2$. It is clearly appreciated that, as the conserved charges associated with the solution become larger, the field gradients become more localized near the boundary region.  As emphasized in \cite{Gentle:2011kv},  the fact that solitons come in one-parameter families with unbounded charge, allows to take a {\it blow up} limit that maps  onto a solution with planar geometry. The procedure starts by looking at the near-boundary expansion
\begin{eqnarray}
&&\phi(t, r) = \frac{\phi_3}{r^3} + O(r^{-4}) \label{phiexp} \\
&&A = A_t (t,r) dt = \left(\mu - \frac{Q}{r}\right)dt + O(r^{-2}) \label{Aexp} \\
&&ds^2 = - \left(r^2 + 1 - \frac{m}{r} \right) dt^2 + \frac{dr^2}{r^2 + 1 -\displaystyle \frac{m}{r}} + r^2 d\Omega_2^2 + O(r^{-2}) \, .\label{gexp}
\end{eqnarray}
Introducing new coordinates
\begin{equation}
r = \lambda \hat{r} ~~~ t = \frac{\hat{t}}{\lambda} ~~~ \theta = \frac{\hat{\theta}}{\lambda} ~~~ \varphi = \hat{\varphi}\, \label{blow_up_map}
\end{equation}
and redefinitions
\begin{equation}
M = \lambda^3 \hat{m}(\lambda) ~~~ Q = \lambda^2 \hat{q}(\lambda) ~~~ \mu = \lambda \hat\mu(\lambda) ~~~ \phi_3 = \lambda^3 \hat{\phi_3}(\lambda) \, ,
\end{equation}
 \eqref{phiexp}-\eqref{gexp} become
 \begin{eqnarray}
 \phi &=& \frac{\hat{\phi}_3}{\hat{r}^3} + O(\hat{r}^{-4}) \label{phiexp2} \\
A &=&  A_t d\hat{t} = \left(\hat{\mu} - \frac{\hat{q}}{\hat{r}}\right)d\hat{t} + O(\hat{r}^{-2}) \label{Aexp2} \\
ds^2 &=& - \left(\hat{r}^2 +  \frac{1}{\lambda^2} - \frac{\hat{m}}{\hat{r}} \right) d\hat{t}^2 + \left(\hat{r}^2 +  \frac{1}{\lambda^2} - \frac{\hat{m}}{\hat{r}} \right)^{-1} d\hat{r}^2 \,  \\
&& + ~ \hat{r}^2 \left(d\hat{\theta}^2 + \lambda^2 \sin\left(\frac{\hat{\theta}}{\lambda}\right)^2 d\hat{\varphi}^2 \right) + O(\hat{r}^{-2}) \, .\label{gexp2}
\end{eqnarray}

\begin{figure}[h!]
\begin{center}
\includegraphics[width=9cm]{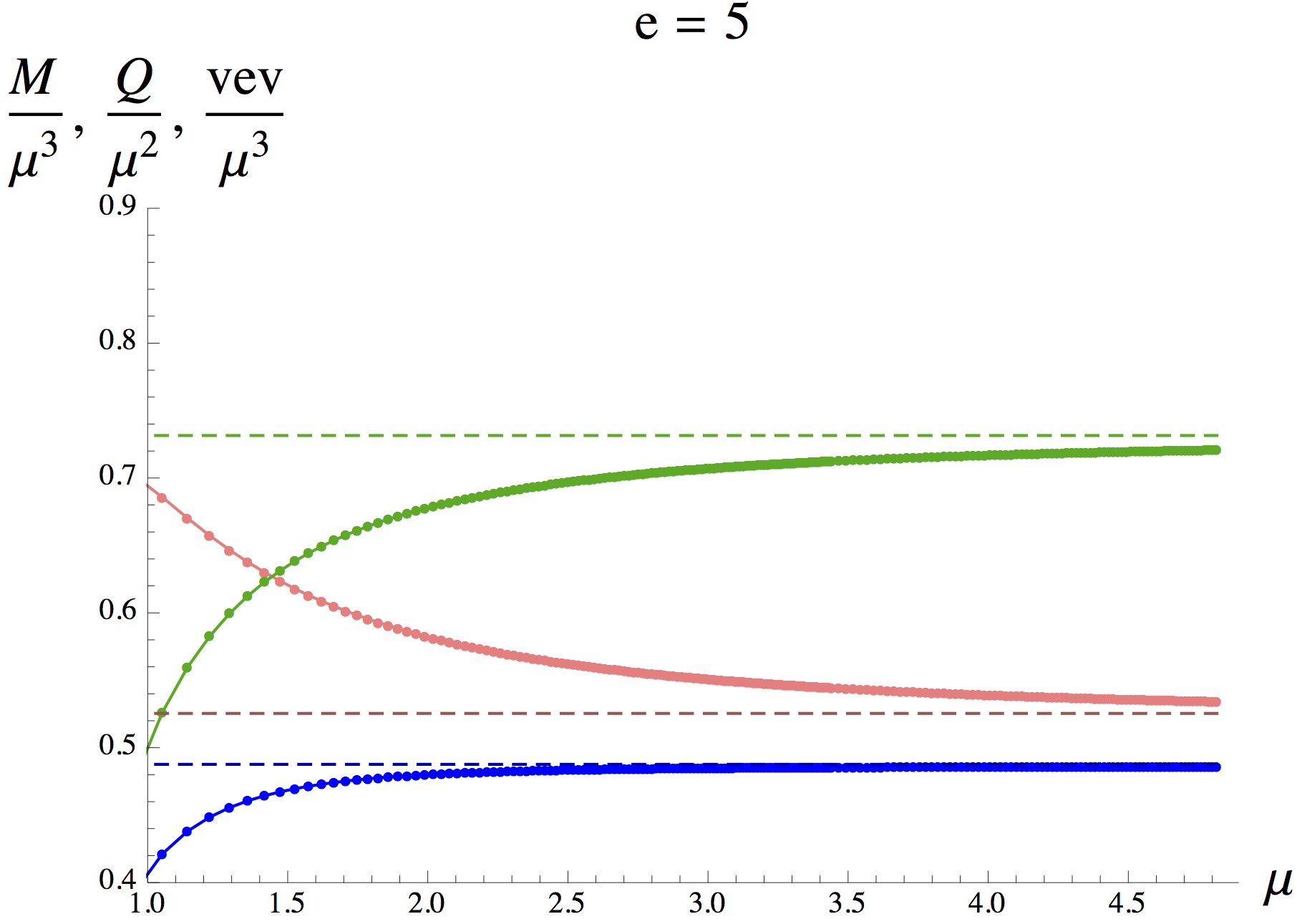}
\caption{\label{scaling_limit} {\small Scale invariant quantities for the $e = 5$ soliton branch. Blue, green and pink colours correspond to the scale invariant ratios of the mass, charge and vev, respectively. Dashed lines correspond to the same ratios computed in the asymptotic planar geometry reached upon blow up.}}
\end{center}
\end{figure}

So far, this is only a reparametrisation of our initial solution. However, if we now take the singular limit $\lambda \rightarrow \infty$ and, simultaneously, move along the soliton branch in such a way that the vector of rescaled quantities $\vec\chi(\lambda) \equiv (\hat{m}(\lambda),\hat{q}(\lambda),\hat{\mu}(\lambda),\hat{\phi_3}(\lambda))$ remains finite, we obtain a planar geometry\footnote{Note that the boundary sphere maps onto a two-dimensional plane in the $\lambda \rightarrow \infty$ limit, with its metric written in polar coordinates.} characterized by the hatted quantities. For example, identifying the parameter $\lambda$ with $\mu$, the dimensionless ratios
\begin{equation}
\frac{M}{\mu^3}, ~~~ \frac{Q}{\mu^2}, ~~~ \frac{\phi_3}{\mu^3} \label{ratios}
\end{equation}
must tend to constant quantities when $\mu \rightarrow \infty$ for the planar limiting geometry to have finite energy, charge and vev densities, respectively. In figure \ref{scaling_limit} the ratios \eqref{ratios} are plotted for the soliton branch at $e=5$. The planar geometry after the blow up limit is taken clearly matches the extremal hairy black brane geometry studied in \cite{Horowitz:2009ij}, at the given $e$.\footnote{We have indeed verified that equations \eqref{static_eqs} reduce to those on \cite{Horowitz:2009ij} after the limit \eqref{blow_up_map} if we set $\kappa^2 = 1/2$ so field normalizations agree.} Incidentally, this observation explains why there does not exist a second soliton branch when $e < e_{t}$. In that case, in planar AdS, there is no near-horizon tachyonic instability that can trigger hair condensation, so there is no limiting extremal solution which this hypothetical branch could map onto.

\section{Grand Canonical ensemble}

The previous section contains a  classification of static solutions without examining the issue about their stability. This is not an unambiguous question, as to compare different solutions one must first specify the ensemble. Whereas the plots we have shown are more akin to studying the system in the microcanonical ensemble, physics is more likely related to the grand canonical ensemble, where temperature and chemical potential are feasible knobs. A very generic feature to take into account is that black hole solutions in global AdS (both RN and hairy) split in two kinds: small and large, having negative and positive specific heat respectively. It is the later ones that should be brought into correspondence with stable superfluid quantum states in the dual field theory. In addition to these thermal solutions, there is the one dual to a thermal gas represented by pure Euclidean AdS$_4$ with a compactified time coordinate. Finally, we have the regular solitons. These static solutions are the building blocks of a rich landscape of first and second order phase transitions.
Along this section we are going to report results for $e=3$. For any coupling $e>e_t$ the situation is qualitatively the same. Below, instead, the transition involves only RN black holes and solitons, since, as evident from figure \ref{menor}, there are no hairy solutions in this regime of coupling. In this case, our system differs from the one analyzed in \cite{Gentle:2011kv}\cite{Basu:2016mol}, where the scalar has a tachyonic mass to start with.

\subsection{Small and large hairy black holes}

\begin{figure}[H]
\begin{center}
\includegraphics[width=7cm]{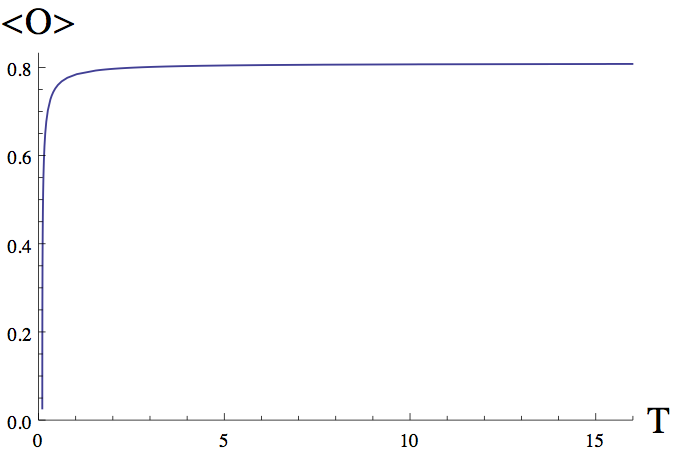}~~~~~~~~~~\includegraphics[width=7cm]{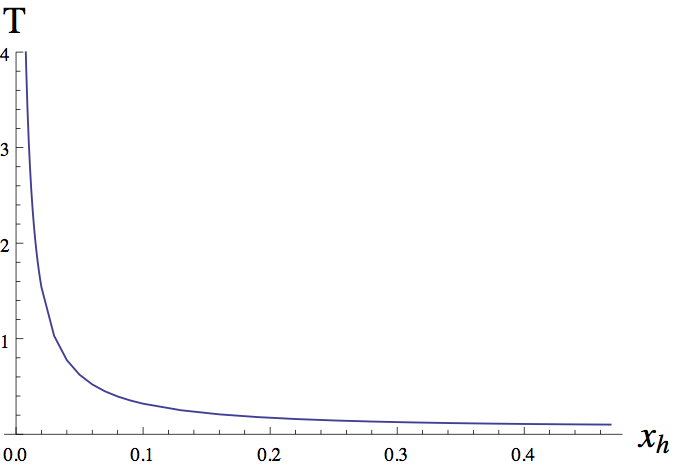}
\caption{\small Left: vev $\langle {\cal O}\rangle$ versus temperature $T$ for $\mu=1.5$ and $e=3$. Right: temperature $T$ as a function of horizon radius $x_h$}
\label{condpi4}
\end{center}
\end{figure}

The thermodynamics of AdS-RN solutions, both in the canonical and grand canonical ensemble has been examined with great care in the past \cite{Chamblin:1999tk}\cite{Chamblin:1999hg}. Here we are going to study the thermodynamical behavior of hairy solutions. The Hawking temperature is given by
\be
T=\frac{e^{\delta(\pi/2)}}{8\pi} \left( \rule{0mm}{4.5mm}(6- A_t'(x_h)^2\cos^4(x_h))\tan(x_h) + 2\cot(x_h)\right)
\ee
and the entropy is
\be
S=2\pi A_h=8\pi^2\tan(x_h)^2, \label{entropia}
\ee
where $A_h$ is the area of the horizon. In figure \ref{condpi4} (left) we plot the vev of the dual operator $\cal O$ for fixed chemical potential $\mu=1.5$ and coupling $e=3$ as a function of the temperature.

The observed behaviour is typical when the gravitational solution corresponds to a small black hole: condensation appears for temperatures greater than some critical value. We will show that this condensed phase is not physically relevant because it has greater free energy than the RN and soliton solutions.
This phenomenon, dubbed retrograde condensation in the literature, has  been observed in different contexts \cite{Jorge}\cite{Cai}. In figure \ref{condpi4} (right) we show the temperature as function of the horizon position for small hairy black holes, where their negative specific heat is manifest.

\begin{figure}[H]
\begin{center}
\includegraphics[width=7cm]{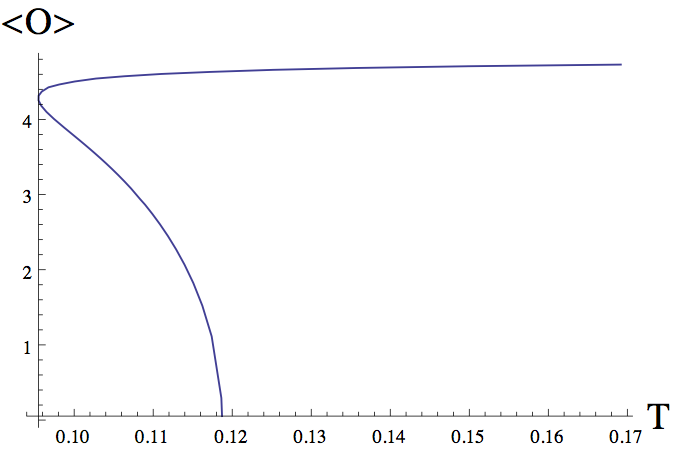}~~~~~~~~~\includegraphics[width=7cm]{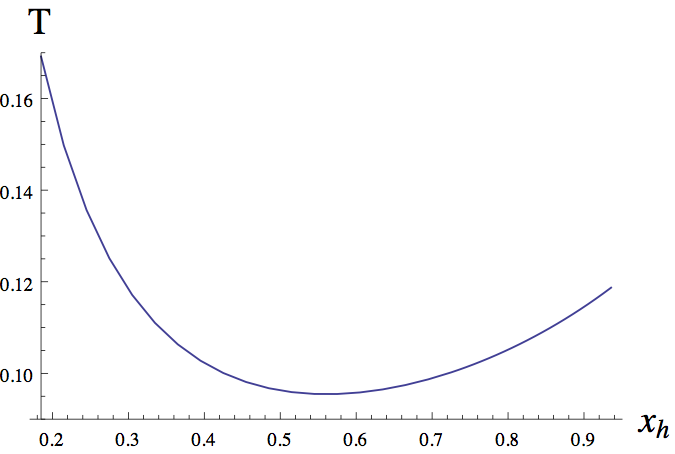}
\caption{\small For $e=3$ and $\mu=3$, on the left, vev $\langle {\cal O}\rangle$ versus the temperature $T$. The lower branch corresponds to the large BH solution and the upper curve to the small BH. On the right, temperature as a function of the position of the horizon. Note that there is one value of $T$ for two different $x_h$.}
\label{fig:e3mu3}
\end{center}
\end{figure}

\begin{figure}[htbp]
\begin{center}
\includegraphics[width=7cm]{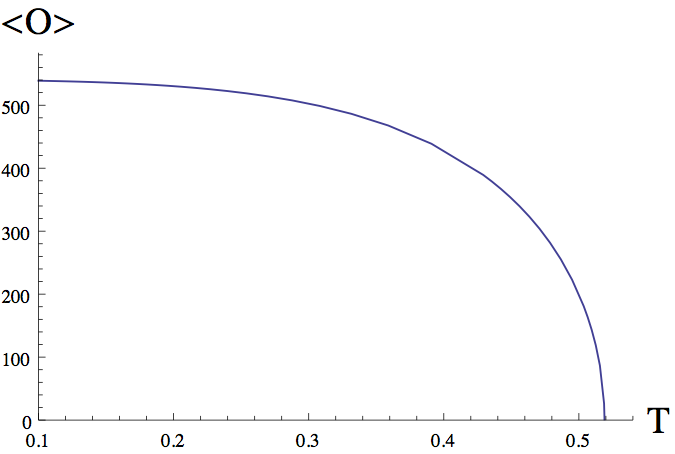}~~~~~~~~~~\includegraphics[width=7cm]{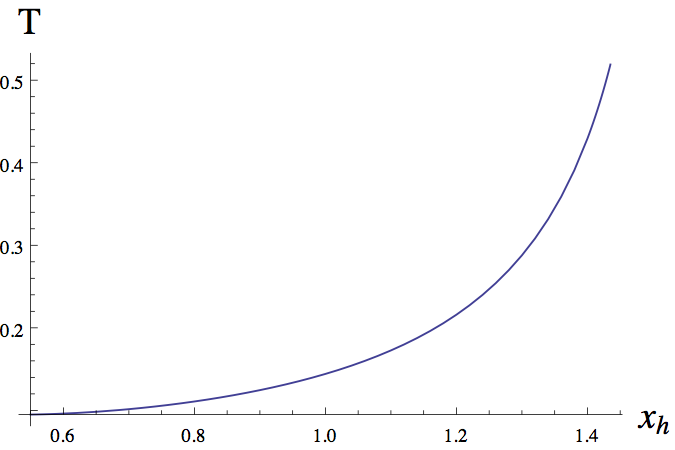}
 \caption{\small For $e=3$ and $\mu=15$, on the left, vev $\langle {\cal O}\rangle$ versus the temperature $T$.  On the right, we observe that the temperature as a function of the horizon position is monotonously growing.}
\label{fig:e3mu15}
\end{center}
\end{figure}

As an example of a small-large black hole transition we rise the value of the chemical potential up to  $ \mu=3$. Figure \ref{fig:e3mu3} (left) shows the behavior of the vev as a function of the temperature. Points on the upper (lower) segment are small (large) hairy black hole solutions. Upon lowering the temperature the system undergoes a second order phase transition along the lower (stable) branch  from a normal to a superconducting state. This can be appreciated in the typical mean field theory behavior of the condensate near the critical temperature, since it goes to zero as $\langle {\cal{O}}\rangle\propto(T_c-T)^{1/2}$. The denomination large/small comes, as usual, from the double valuedness of the temperature as a function of the horizon radius $x_h$. This is seen on the right plot of figure \ref{fig:e3mu3}.
Further increasing the chemical potential, we observe the behaviour shown in figure \ref{fig:e3mu15} for $\mu=15$. The same  second order phase transition
happens here, but the small black hole branch has now disappeared.

\subsection{Grand potential}

Linear stability  of the previous solutions amounts to the minimisation of certain thermodynamical potential that depends on the ensemble. In the grand canonical ensemble this is the grand potential $\Omega(T,\mu)$, which the AdS/CFT correspondence identifies  with the on-shell renormalized Euclidean action $\Omega = T S_{\rm on-shell}$  (see appendix C for calculations).

\begin{figure}[h]
\begin{center}
\includegraphics[width=7.5 cm]{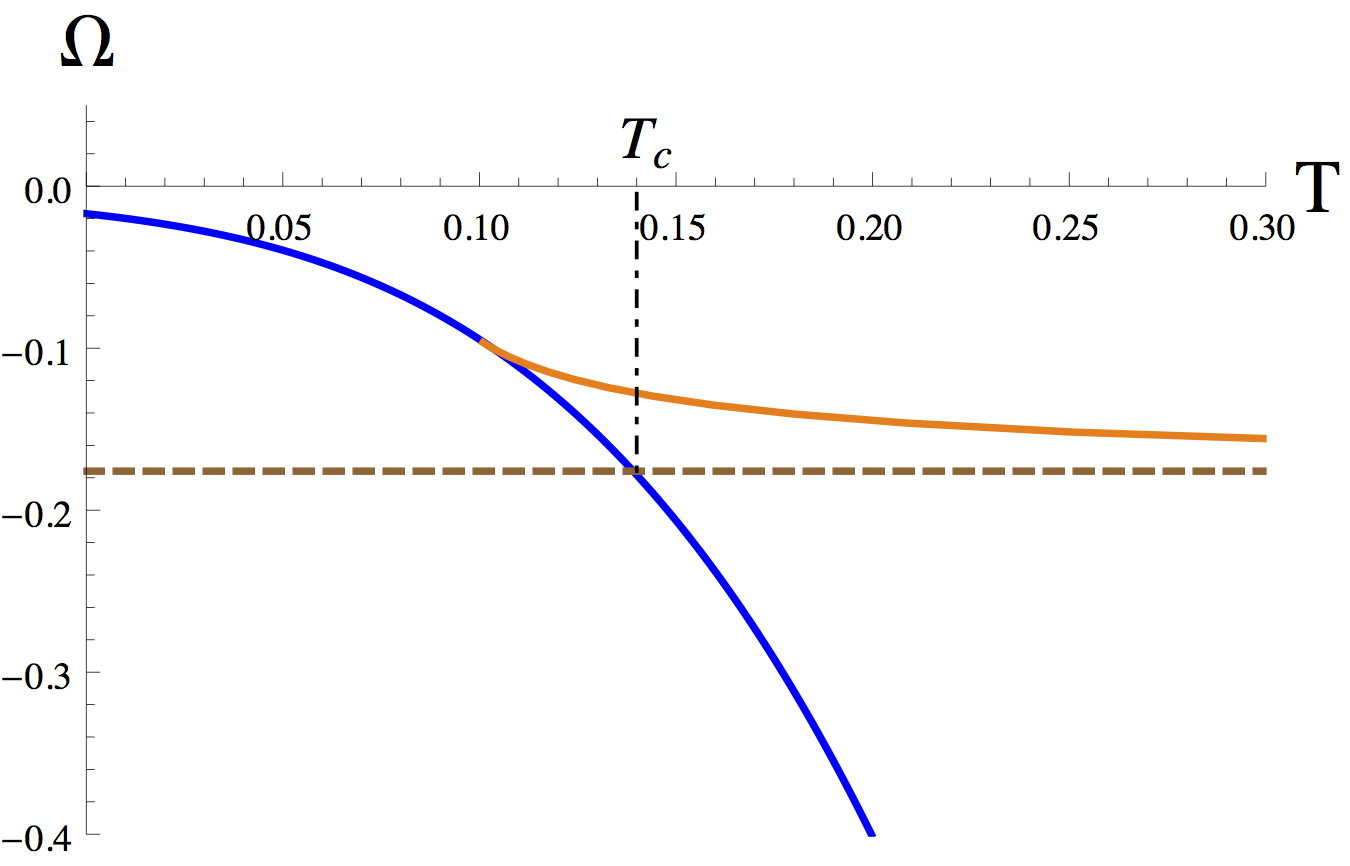}~~~~~~~~~\includegraphics[width=7.5cm]{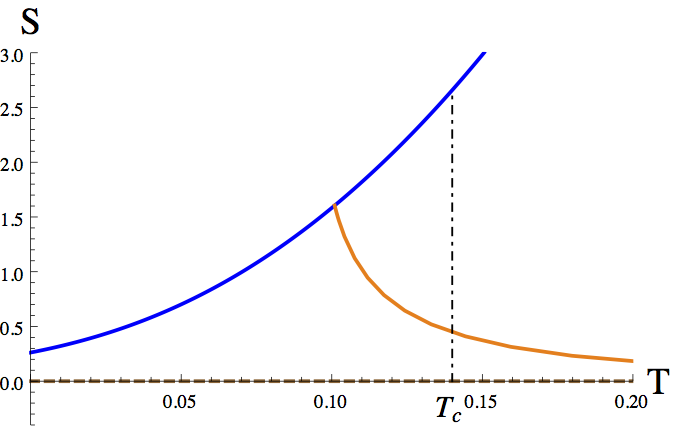}
\caption{\small $\mu=1.5$, $e=3$. The thermodynamic potential (left) and entropy (right) as function of the temperature. The blue line corresponds to the RN solution, the orange curve is for the small hairy black hole solution and the brown dashed line stands for the soliton. There is a critical temperature, $T_c$, below which the soliton is the dominant solution. We observe that the entropy is discontinuous at the meeting point between the brown and blue curve showing the appearance of a first order phase transition at $T_c$.}
\label{Free}
\end{center}
\end{figure}
In figure \ref{Free} we plot the grand potential \eqref{potential} and the entropy \eqref{entropia} respectively as functions of the temperature. In both figures the blue curve refers to the RN solution and the orange curve to the small hairy black holes for $\mu=1.5$ and $e=3$.
\begin{figure}[H]
\begin{center}
\includegraphics[width=7.5cm]{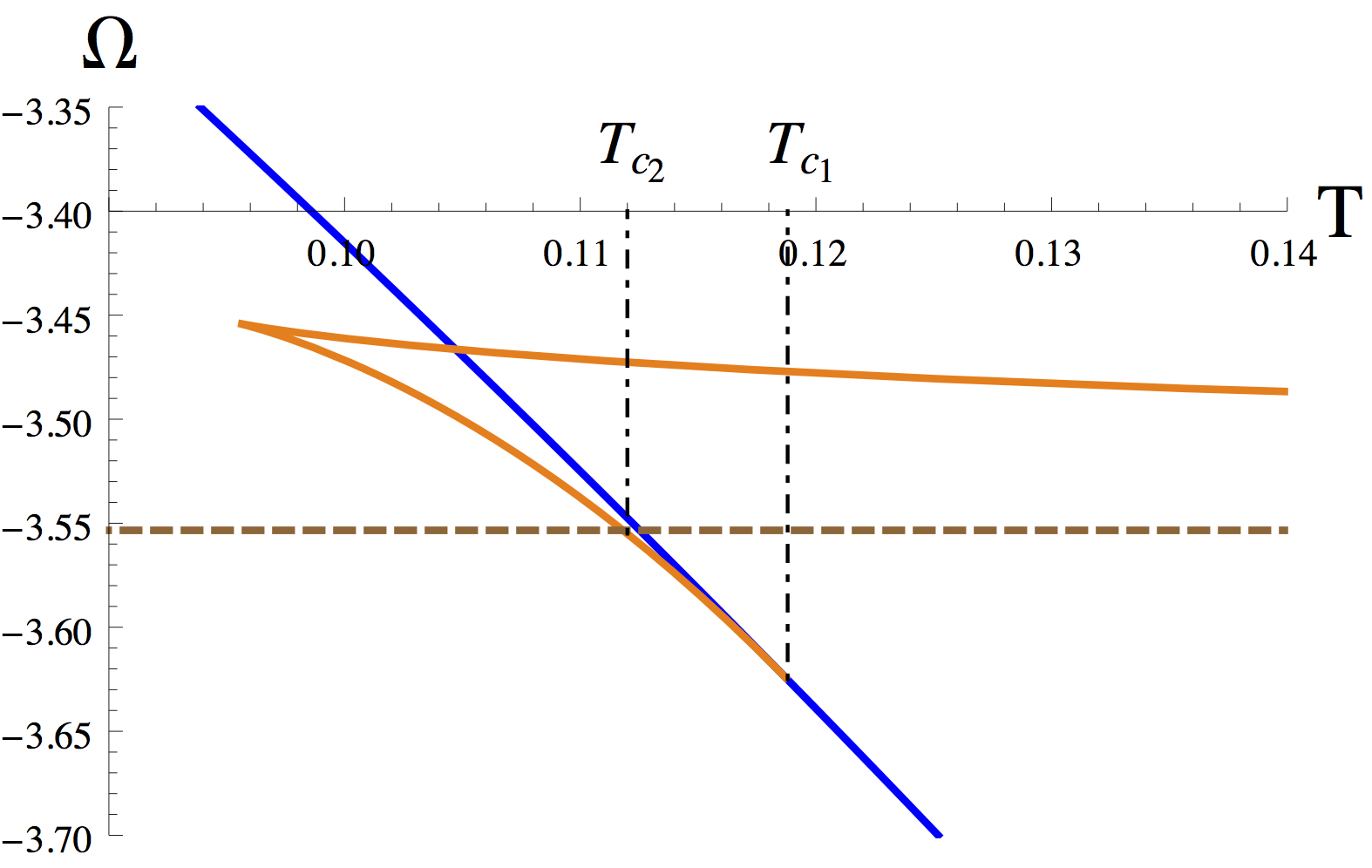} ~~~~~\includegraphics[width=7.5cm]{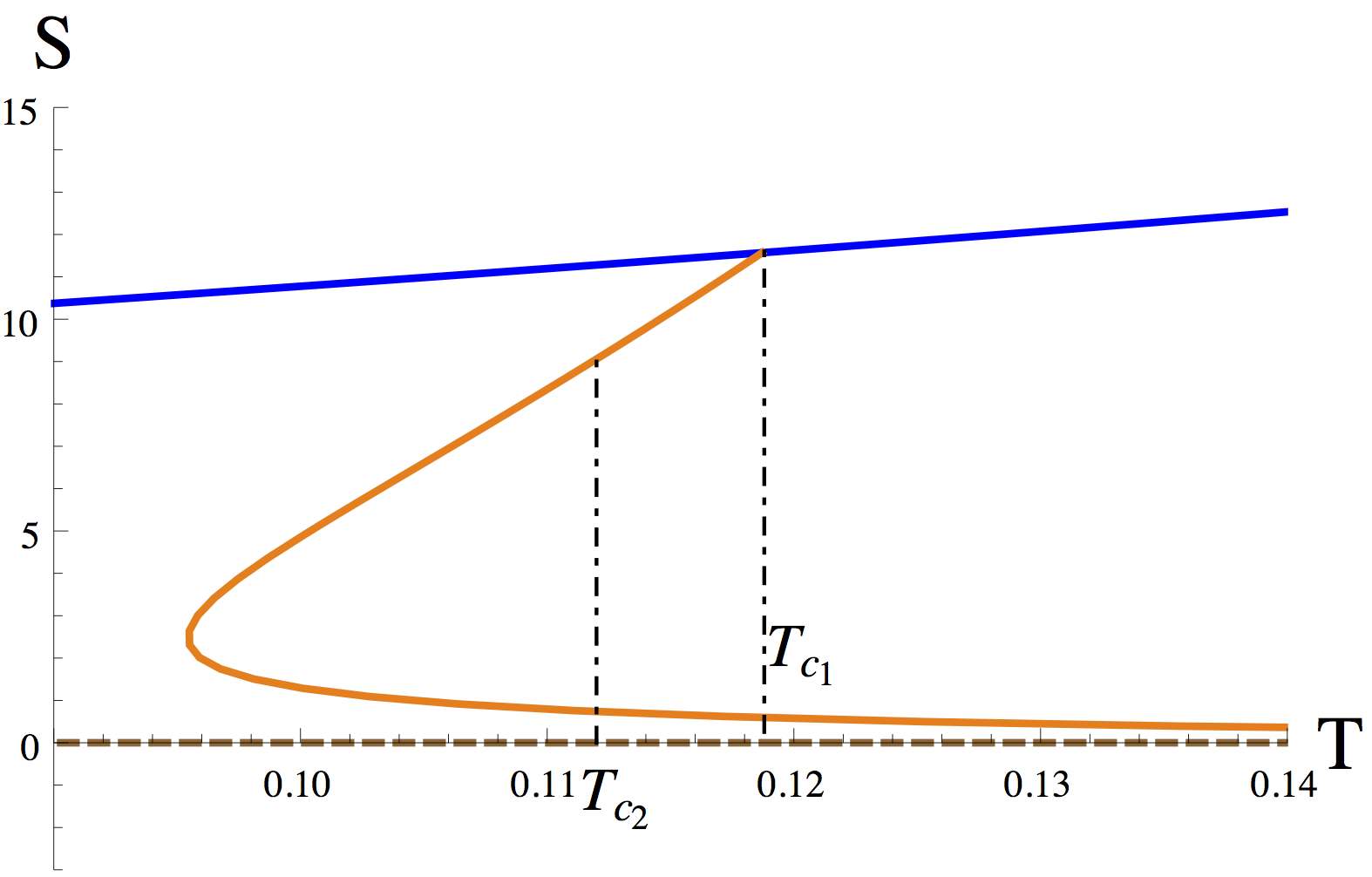}
\caption{\small $\mu=3$, $e=3$.  The thermodynamic potential (left) and entropy (right) as function of the temperature. The blue line corresponds to the RN solution and the orange curve to the hairy black hole solution. The hairy solution has two branches, the lower one corresponds to the large BH and the upper one to the small BH. The brown dashed curve denotes the soliton free energy.}
\label{Freesmall}
\end{center}
\end{figure}
\begin{figure}[htbp]
\begin{center}
\includegraphics[width=7.3cm]{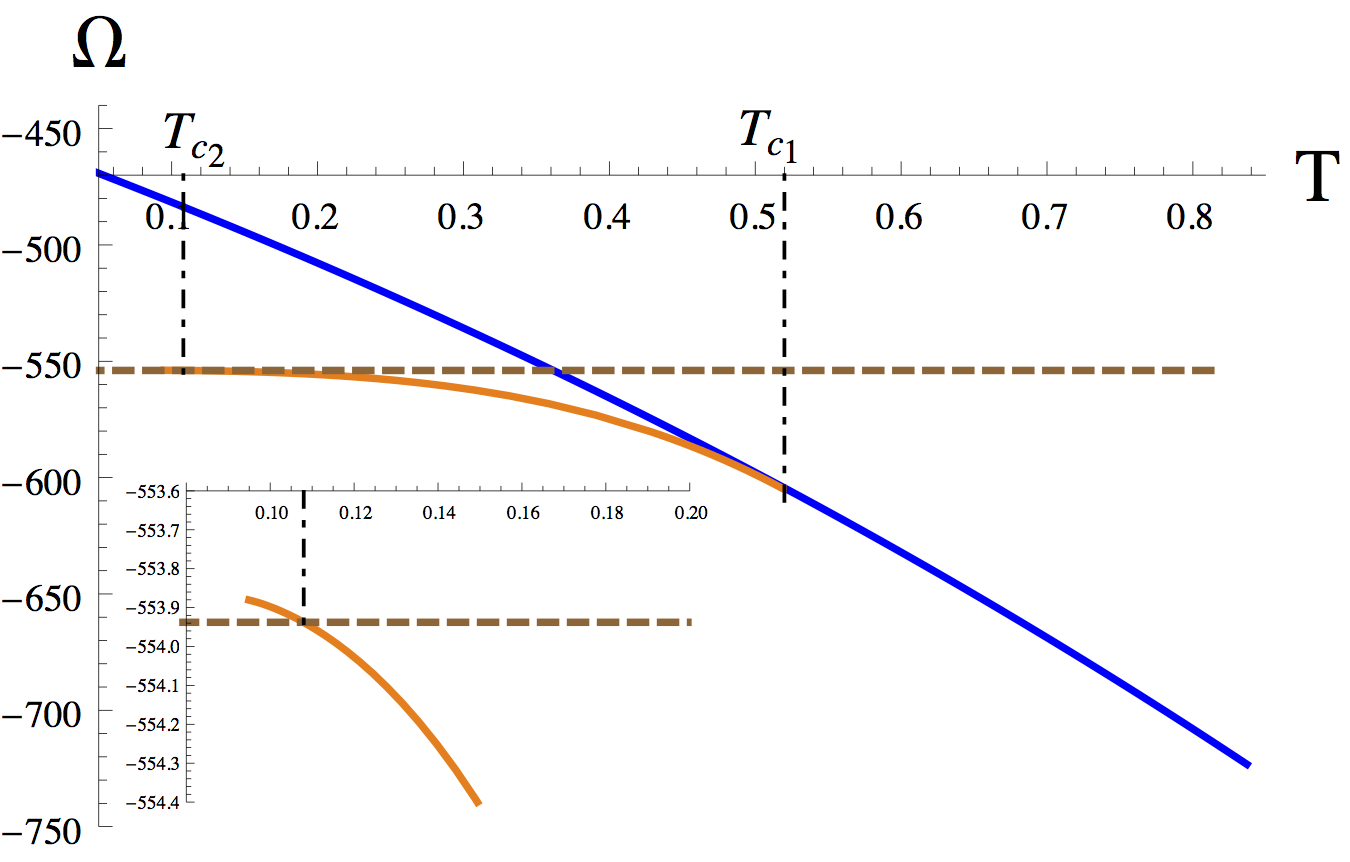} ~~~~~~\includegraphics[width=7.5cm]{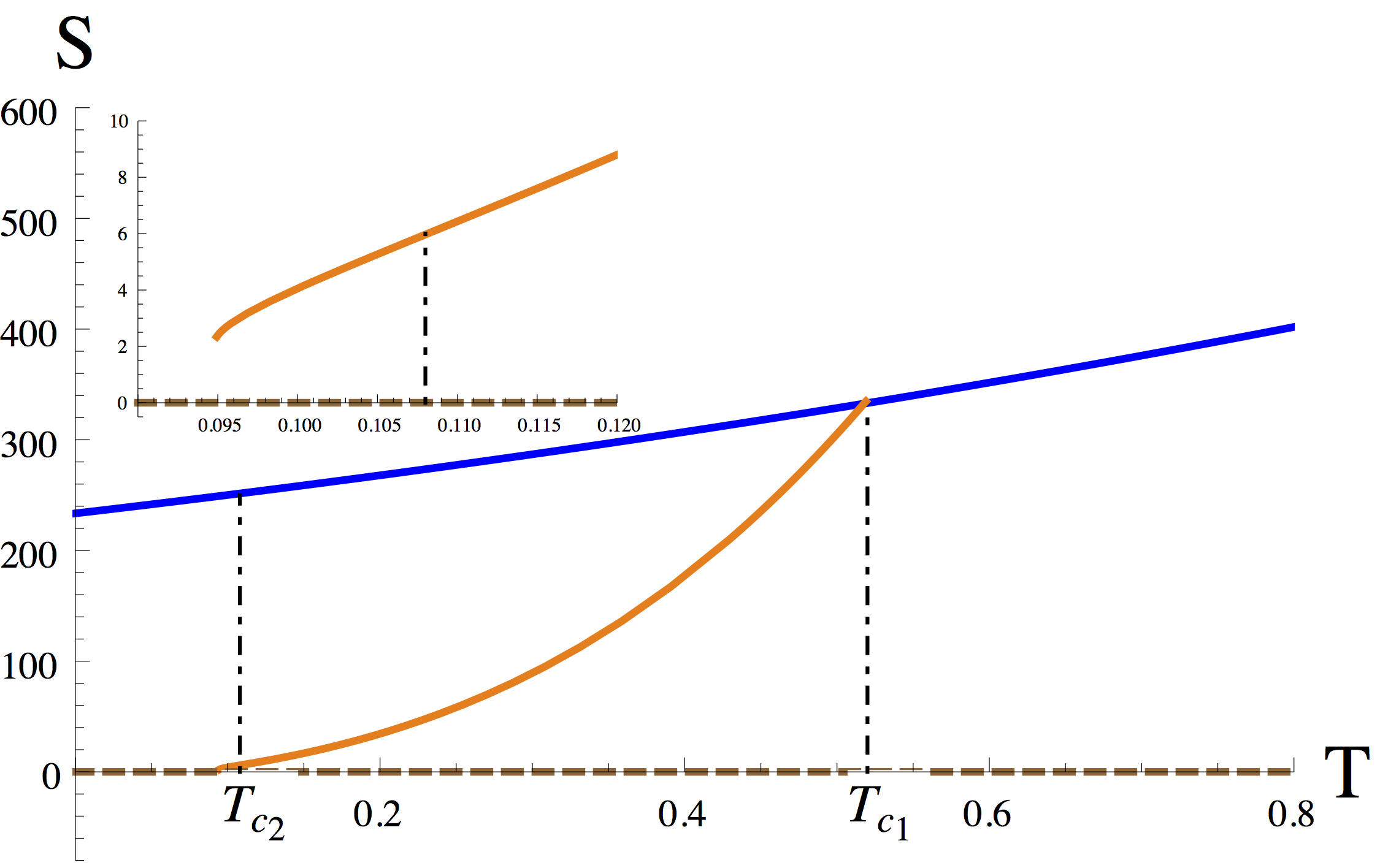}
\caption{\small $\mu=15$, $e=3$. Thermodynamic potential (left) and entropy (right) as function of the temperature with the same color coding as in the previous figures.  There are no small hairy black holes in this regime.}
\label{Freemu15}
\end{center}
\end{figure}

The brown dashed line is the free energy of the soliton.\footnote{Since the soliton geometry is regular at $x = 0$, it can be analytically continued to a smooth Euclidean geometry with any period in the time direction. Choosing a finite period corresponds to placing a thermal gas with $O(1)$ entropy and energy above the soliton geometry. A natural way of fixing this period would be to demand continuity of the isotherms in the $(Q, M)$ phase diagram, which would require to assign an infinity temperature to the unbounded soliton branches. However, since they correspond to zero entropy ground states, another natural criterium would be to demand a vanishing temperature. This choice is also consistent with the fact that, in the blow up limit discussed in section 2, they map onto extremal hairy black branes.} We observe that a first order phase transition occurs, whereby increasing $T$ the solitons switch into RN black holes. The small hairy black holes are never the ground state at this value of  $\mu$.
Figure \ref{Freesmall} shows the same functions for the $\mu=3, e=3$ solutions. The  preferred branch is that of  large black holes (lower orange segment) instead of small black ones (upper orange line) which have negative specific heat (see figure \ref{fig:e3mu3}). Again, the derivative of the entropy is discontinuous at a critical temperature $T_{c_1}$ and we have a second order phase transition denoting the normal-superconductor transition of the dual QFT. We detect  a second first order phase transition temperature $T_{c_2}$, below which solitons are preferred.

Finally, figure \ref{Freemu15}  corresponds to $\mu=15$. The phase space is the same, as well as the nature of the phase transitions, the only difference being  the disappearance of the small black hole solutions.

\section{AdS nonlinear stability}

The above sections have relied on a combination of analytical and numerical arguments. The  construction of static solutions is  performed by solving ODE's and setting up a shooting procedure.
It involves fixing, for example, the radius of the desired solution and the value of the scalar field $\phi_o$ at such radius. Then the value of $A_t'(x_o)$ is also varied until one obtains
a solution with vanishing source $\phi_b = 0$ and non vanishing vev $\phi_{b,3}\neq 0$.  The end result is scrutinized to find the actual value of the mass $M$ and the charge $Q$ of the obtained stationary solution. If more than one solution is available, the free energy is to be invoked in order to select the correct ground state.
In a sense, this strategy relies on a certain amount of guesswork. Prior to the construction of hairy black holes in  \cite{Gubser:2008px},  the space of known static vacua consisted of either pure AdS or  RN black holes. Later on, solitons where first inferred, and then constructed  \cite{Basu:2010uz} from a limit whereby the hairy black hole's horizon is shrunk to zero.

In this section we will use a complementary approach.  A numerical simulation code for time evolution is the closest thing one can have to a real experiment. In this spirit, the approach starts from the other end: one devises a certain initial radial profile for the bulk fields, with a given total mass $M$ and charge $Q$, and lets it evolve under a
scheme that preserves these values.\footnote{The precise implementation of the time evolution code, as well as its convergence properties, are discussed in appendix D.}
If the evolution settles down to a certain stationary state, it must necessarily be one in the list  above. And if two of them are available with the same values of $M$ and $Q$,  the evolution will select the ground state in the microcanonical ensemble.\footnote{This could proceed in a direct way or through a number of different steps.  There are situations where pre-thermalization to some excited intermediate state followed by further relaxation to the true final equilibrium state can be observed, for instance in the latest paper of \cite{Gursoy:2016ggq}. It would be interesting to study in detail the possible existence of such metastable attractors in the present context.}

By evolving an initial condition below the blue curve in figure \ref{fig2} this would have
shown that  black holes with abelian hair exist in global AdS, had this paper been written prior to 2008. Moreover, imagine there was any other exotic type of black hole that nobody has constructed yet using static methods (ODE) and, moreover, suppose  it has larger entropy. The collapse simulation would {\em smell} its existence and  the fields  decay to that solution after exploring large portions of phase space. We must admit we have not found any new such solution  using this, admittedly expensive, method.

\begin{figure}[htbp]
\begin{center}
\includegraphics[scale=0.62]{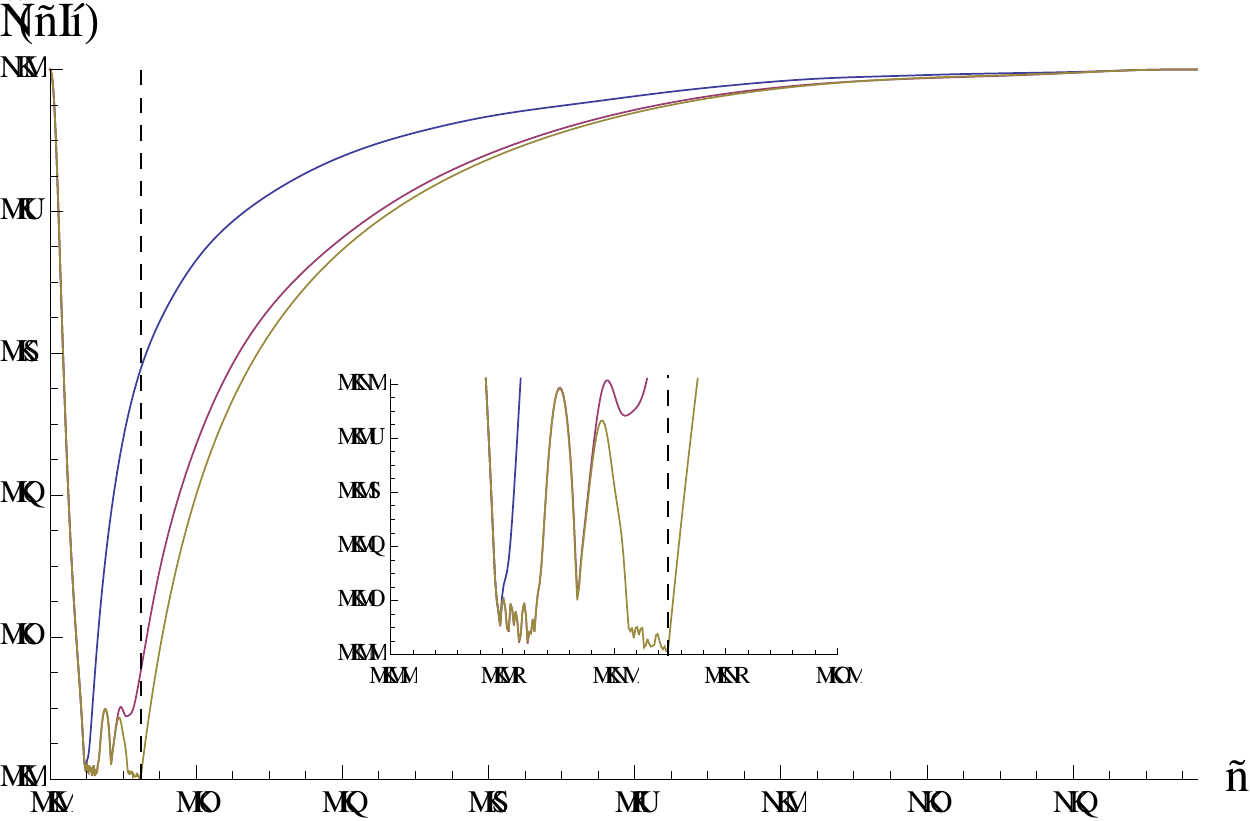} ~~~~\includegraphics[scale=0.62]{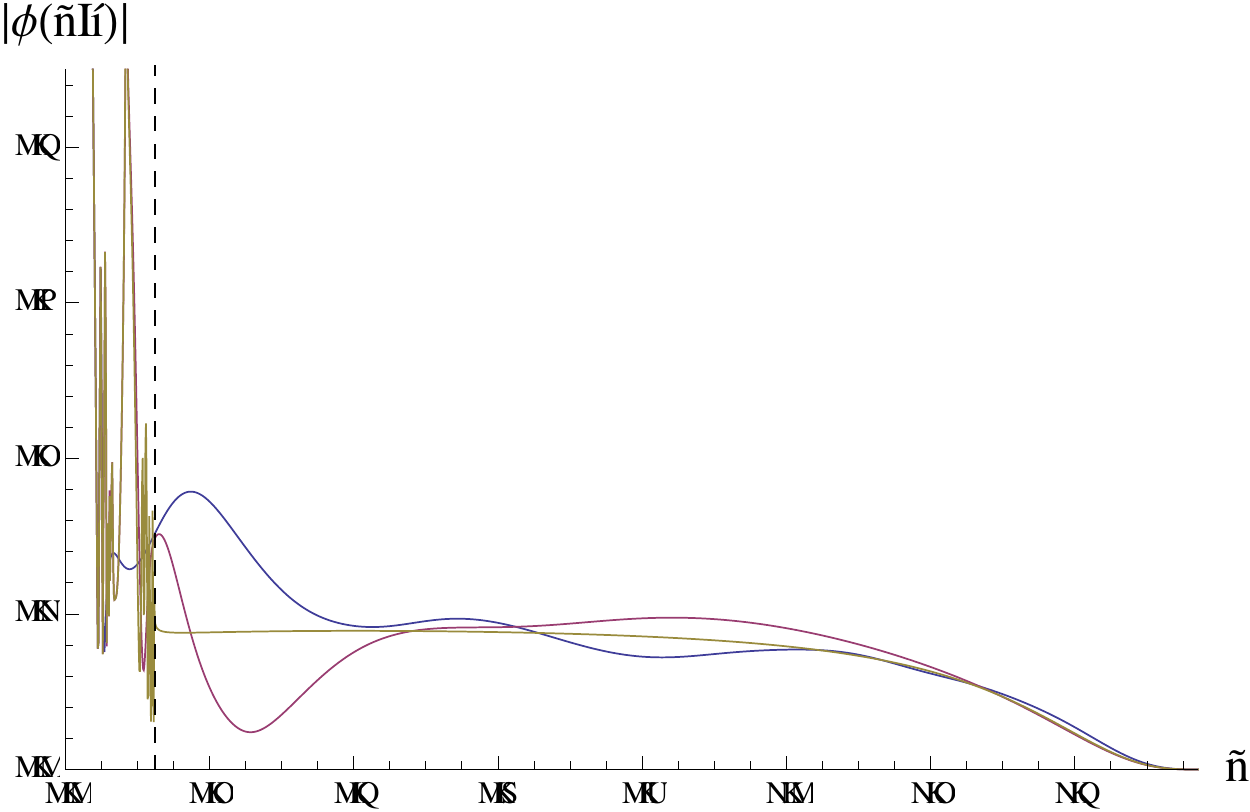}
\caption{\small  For  coupling $e=20$, we plot three snapshots of a collapse with mass $M=2$ and charge $Q=2.9$ at times $t=14.8 , 17.3$ and $28.1$ (blue, magenta and yellow). At this time, solution has become an almost static hairy black hole. The evolution exhibits the typical ring down whereby apparent horizon builds up while capturing successive oscillations of the scalar field. They fold behind the accumulation point of the vanishing locus of $f(t,x)$, hence the position of the event horizon  }
\label{fig:hairydecay}
\end{center}
\end{figure}

In figure \ref{fig:hairydecay} we have plotted some snapshots of a typical collapse to a hairy black hole. For the values of charge, mass, and coupling $e$ used in this simulation there is no RN black hole available. Nevertheless at $t=14.8$ the scalar develops a spike
at a point where the metric approaches an apparent horizon, signalled by a zero of the function $f$ (blue curve). Even if the zero value is never reached, the dynamics close to this point becomes extremely slowed down in terms of the time at the boundary $x=\pi/2$.
At later times, the outer oscillations of the scalar field start piling up on top of the first spike, and the metric function $f$ tries to reach zero at higher values of the coordinate $x$ (see inset). A very high precision and up to $2^{17}$ grid points are needed to push this
numerical evolution safely, and resolve the region close to the collapse with enough accuracy, in particular monitoring the constancy of $M$ and $Q$ values throughout the process.  The exponentially decaying ringdown ends up in a static solution where the outside hair profile
resembles the ones in figure \ref{fig2} for the soliton solutions.

In ref. \cite{Dias:2012tq} a strong claim was made, that the  geometries exhibiting a {\em corner of instability}, were those for which the spectrum of linear perturbations is fully resonant. An important case that shows how sensitive the dynamics is to the resonance property can be found in the case of a scalar field on asymptotically flat space enclosed in a cavity. An instability corner is obtained for Dirichlet boundary conditions \cite{Maliborski:2012gx}, where the spectrum is resonant, but not instead in the case of Neumann boundary conditions \cite{Maliborski:2014rma}, where it is only asymptotically resonant.\footnote{See also  \cite{Okawa:2014nea}} The upshot of these analysis seems to indicate that, not only an UV cascade is important, but also a non dispersive spectrum, such that the initial wave packet remains focused at all times.

To test this picture further, one would be interested in families of situations that depart smoothly from it and, in this sense, there are two ways to achieve this. One can keep the original lagrangian setup and, say, consider perturbations around equilibrium solutions other than pure AdS. In generic cases, the spectrum of perturbations of the scalar laplacian on this new background will not be resonant and the collapse will not exhibit a corner of instability. This is the proposed mechanism for the stability of broad initial data, as studied in \cite{Maliborski:2014rma}. They can be understood as perturbations of a single oscillon solutions which are known to be stable. A different strategy is to deform the theory by a continuous parameter. Examples of this are Gauss Bonnet \cite{Deppe:2014oua}, or the inclusion of a hard wall \cite{Craps:2014eba}. A remarkable example that stands out is AdS$_3$, which despite being resonant, has a mass gap \cite{Bizon:2013xha} \cite{daSilva:2014zva}.

The present setup offers another interesting example where the theory is deformed by the presence of additional degrees of freedom, while still preserving the resonant spectrum.
At the linear level, scalar perturbations obey the same equations of motion as
in Einstein-scalar theory.\footnote{We thank Oscar Dias for this observation.}  A perturbation of the (now charged) scalar field of  amplitude $\epsilon$ sources a gauge field $A_\mu$ at the same order,
which will  backreact on the scalar equation of motion at next order $\epsilon^2$ since it couples through the covariant derivatives. Hence we have another work bench to test whether nonlinear perturbations exhibit a corner of instability.

In ref. \cite{Buchel:2012uh} collapse of  a complex scalar was considered and no significant difference was obtained for the phenomenology of charged vs. neutral configurations, apart from a small decrease in collapse time.
The results in that paper can be  recovered in the limit of vanishing coupling $e\to 0$ of our work. However, at finite $e$ we find  opposite results as we will now see.

\subsection{Initial conditions}

In this section we will show the results of uniparametric families of collapses that scan across the $(Q,M)$ plane.  The first protocol will involve a set of initial conditions, parameterized with some amplitude $\epsilon $.
 Consider the following family of gaussian initial data that initially fall from the boundary
$\Phi= \Phi_1 + i\Phi_2$, and $\Pi = \Pi_1 + i\Pi_2$ with $\Phi_2= \Pi_1 = 0$ and
\beqa
\Phi_1 &=& \epsilon\cos \beta\, \frac{2}{\pi} \cos^{2} x \exp\left( -\frac{4 \cot^2(x)}{\pi^2 \sigma^2} \right) \label{gauss1} \\
\Pi_2 &=& \epsilon \sin\beta\,  \frac{2}{\pi} \cos^3 x \exp \left( -\frac{4\cot^2(x)}{\pi^2 \sigma^2}  \right) \, . \label{gauss2}
\eeqa
\begin{figure}[htbp]
\begin{center}
\includegraphics[width=7.8cm]{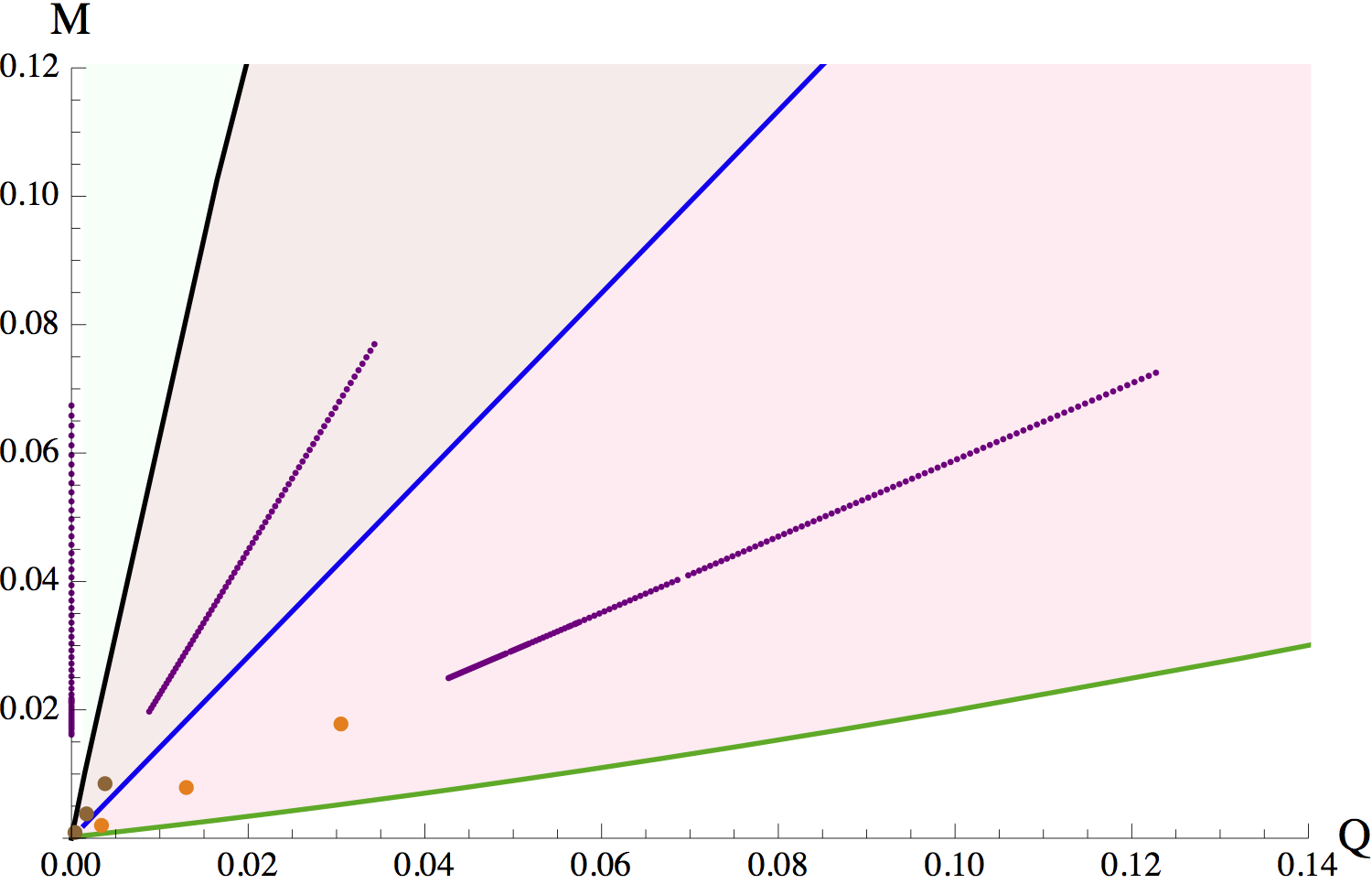}~~~~~~~\includegraphics[width=7.5cm]{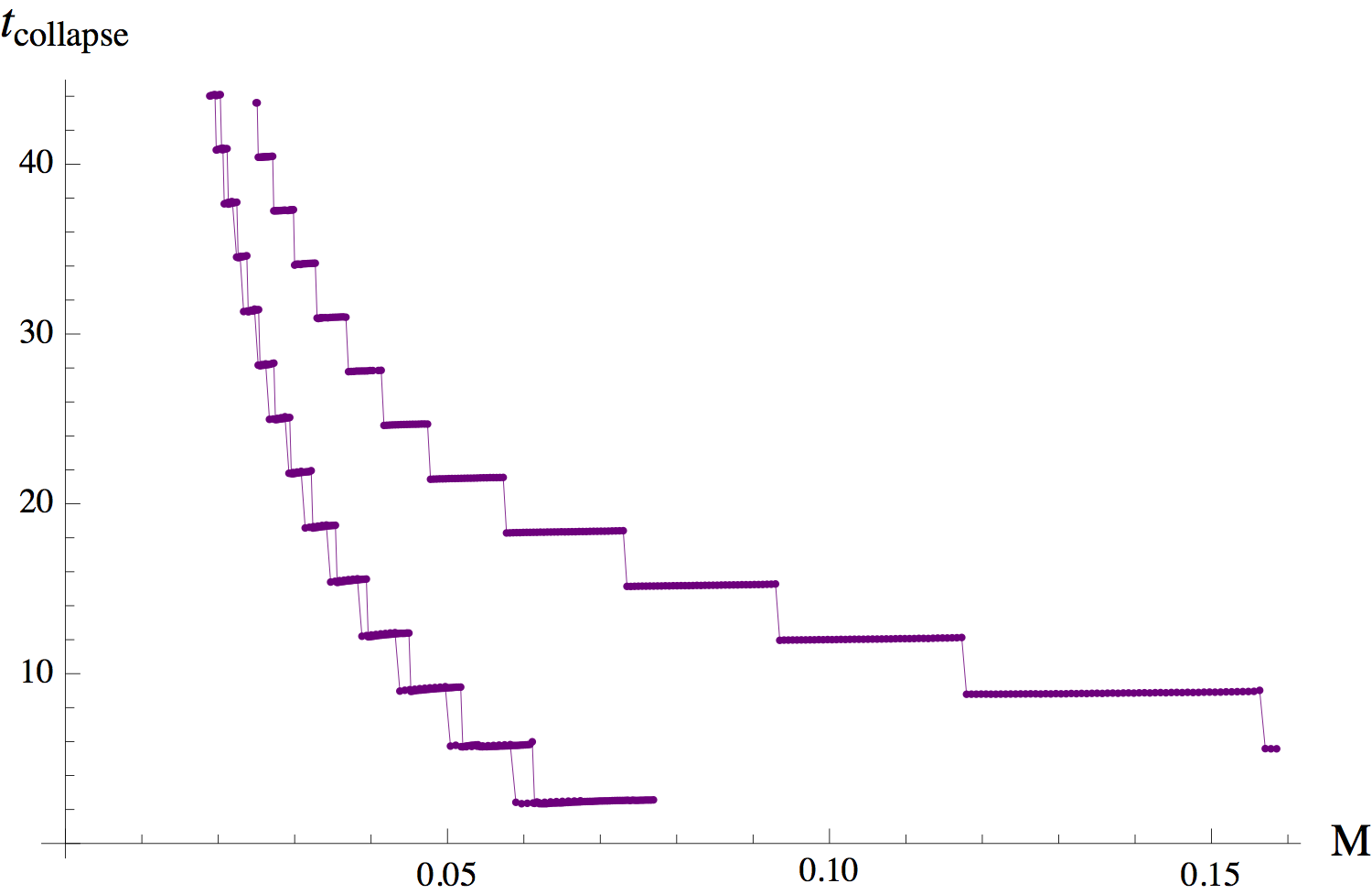}
\caption{\small  Collapses with $\sigma=0.1$ and $e=20$. In the left plot, the phase space and color coding is the same as in figure \ref{fig1}. The dotted purple lines represent three series of initial conditions with $\beta=0^\circ, \, 45^\circ$ and $82^\circ$ from left (vertical) to right (most bended).
 For each line we have taken 3 points  deep in the low $\epsilon$ limit with $\epsilon=6,4$ and $2$, and performed very long time simulations  in order to test the scaling hypothesis (see figure \ref{fig:imecollapsesigma0d2e20}). Right plot: time for collapse for each of the three lines in the left plot, in the same order from left to right, as a function of the mass $M$. The time for collapse increases with the closeness to the soliton line, i.e. for fixed mass $M$, at higher charge $Q$. }
\label{fig:imecollapsesigma0d1e20}
\end{center}
\end{figure}
Here the angle  $\beta$ is fixed, and  $\epsilon$ will decrease monotonically towards zero.  The cases $\beta=0,\pi/2$  corresponds to the uncharged initial conditions studied in \cite{Bizon:2011gg}, albeit with an initial pulse that starts infalling from the boundary \cite{Abajo-Arrastia:2014fma}, a fact that is inspired from the physics of a quench. For $\beta \neq 0,\pi/2$  we are dealing with a  shell of charged scalar and gauge field collapsing together.
\begin{figure}[H]
\begin{center}
\includegraphics[width=8cm]{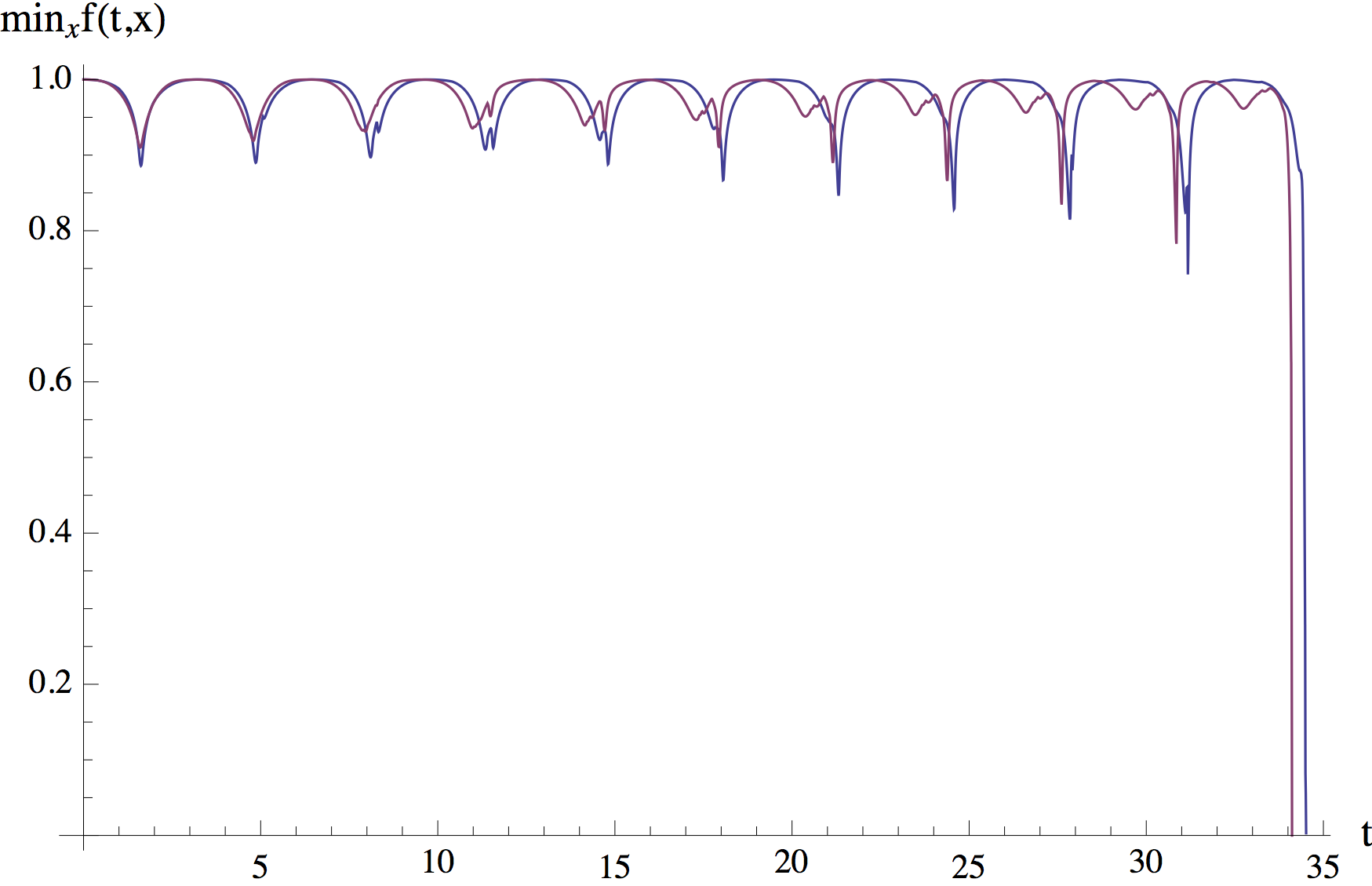}
\caption{\small \label{fig:minf} Plot of the minimum of the metric function $f(t,x)$ as a function of time for two simulations with similar histories in the lines $\beta=45^\circ$ (blue) and $\beta=82^\circ$ (magenta) in figure \ref{fig:imecollapsesigma0d1e20} left. The roughness of the
profile in the second case reveals that the scalar pulse is not smooth but fractures into ripples. This, presumably, is an effect of the electrostatic repulsion in action.}
\end{center}
\end{figure}

In  figures \ref{fig:imecollapsesigma0d1e20} and \ref{fig:imecollapsesigma0d2e20} we exhibit series of simulations for initial conditions of the form \eqref{gauss1}\eqref{gauss2} $\sigma=0.1$ and $0.2$, and several values of $\beta$. The cases examined in  \cite{Bizon:2011gg} would lie on the vertical axis $Q=0$ (uncharged case).
Notice that we have searched for peculiar behaviours in different domains of the phase space in this microcanonical ensemble.  Above the instability line where only RN solutions exist, we find little or no difference with the case of zero charge.   Below this line, the final state of the evolution is a hairy black hole.
We have included two lines of collapse that  ``bend'' towards  the soliton line (the green line), in the region where hairy black holes exist, both above and below the  line of  extremal RN (in blue).

\begin{figure}[htbp]
\begin{center}
\includegraphics[scale=1.0]{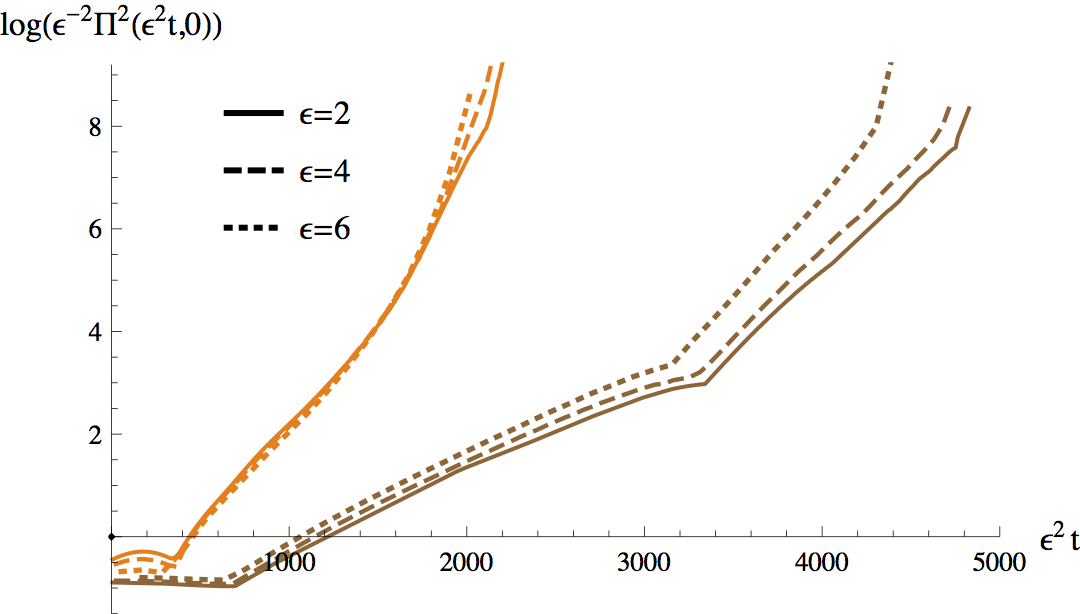}
\caption{\small Evolution of the maxima of the Ricci scalar  at the origin upon rescaling of the time and the initial amplitude. The six simulations correspond to the six dots in figure \ref{fig:imecollapsesigma0d1e20} with $\sigma=0.1$ and
$\beta=45^\circ$ (brown) and $82^\circ$ (orange).}
\label{fig:xscalingplot}
\end{center}
\end{figure}
We want to stress that it is by no means easy to engineer initial conditions that come close to the line of soliton solutions.  In particular, within the family of gaussians spelled out in \eqref{gauss1} and \eqref{gauss2}, by letting $\beta$ sweep from 0 to $\pi/2$, the lines incline up to some point, for some $\beta_0$, where the initial conditions approaches maximally the soliton line and then turn back towards the vertical. These values are, for example,  $\beta_{0} = 82^\circ$ for $\sigma=0.1$ and $\beta_0 = 75^\circ$ for $\sigma = 0.2$. In principle one can engineer initial conditions that come closer to the soliton line by starting from the other end: namely, by perturbing a soliton, and this will be the subject of the next  section.
Most remarkable is the fact that is seems impossible to even write
initial data whose charge, $Q$,  and mass, $M$, give a point below the soliton line (white region). This seems to point out that soliton solutions extremize certain positive definite functional that can be derived from the action. It would be very interesting
to elaborate on this point further.

Time for collapse is one of the important observables in the game. We can see in plot \ref{fig:imecollapsesigma0d1e20} the case of sharp pulses with  $\sigma=0.1$ and $\beta=0^\circ,\,48^\circ$ and $82^\circ$. The horizontal axis represents the initial
mass, $M$, which, for fixed $\sigma$, grows with  $\epsilon^2$.   On the right, the times for collapse for each family are plotted.  When moving along an individual series from right to left, the plateaux reflect the number of oscillations that the system undergoes before the final collapse is reached. The fact that electromagnetic repulsion counteracts gravitational attraction is probably behind the fact that oscillations exist at higher values of  $M$ when the charge is larger.   We see that the behaviour points towards the existence of a corner of instability at the origin even  in the charged situation, i.e., no sign of a threshold for stability is appreciated.
This seems to confirm the expectation coming from the resonant character of the linearized approximation.
From the right  figure \ref{fig:imecollapsesigma0d1e20} we draw the important conclusion that charged configurations take longer time to collapse. This is in  sharp  contrast with the case without gauge field where charged initial conditions
were collapsing sooner than neutral ones of the same mass (see figure 6 in \cite{Buchel:2012uh}).

In the present situation, the charge of the pulse also adds to the defocusing. Still, what figure \ref{fig:imecollapsesigma0d1e20} says is that, for $\sigma =0.1$, this is not enough to
erase the instability corner, even for the most charged gaussians that one can device (the right most blue magenta diagonal). Both plots for collapse time scale with $1/M \sim 1/\epsilon^2$
in the limit $\epsilon\to 0$. However, a closer look at the evolution of the scalar field reveals that, for $\beta = 89^\circ$, the initial gaussian develops subpulses. A reflection of this can be observed in figure \ref{fig:minf}  which plots
the minimum of the metric function min$_xf(t,x)$ as a function of time.
The evolution  still exhibits a quasiperiodic structure where the action of the weak turbulent cascade is apparent in that the minima become sharper and deeper until finally collapse takes over.

\begin{figure}[htbp]
\begin{center}
\includegraphics[scale=0.62]{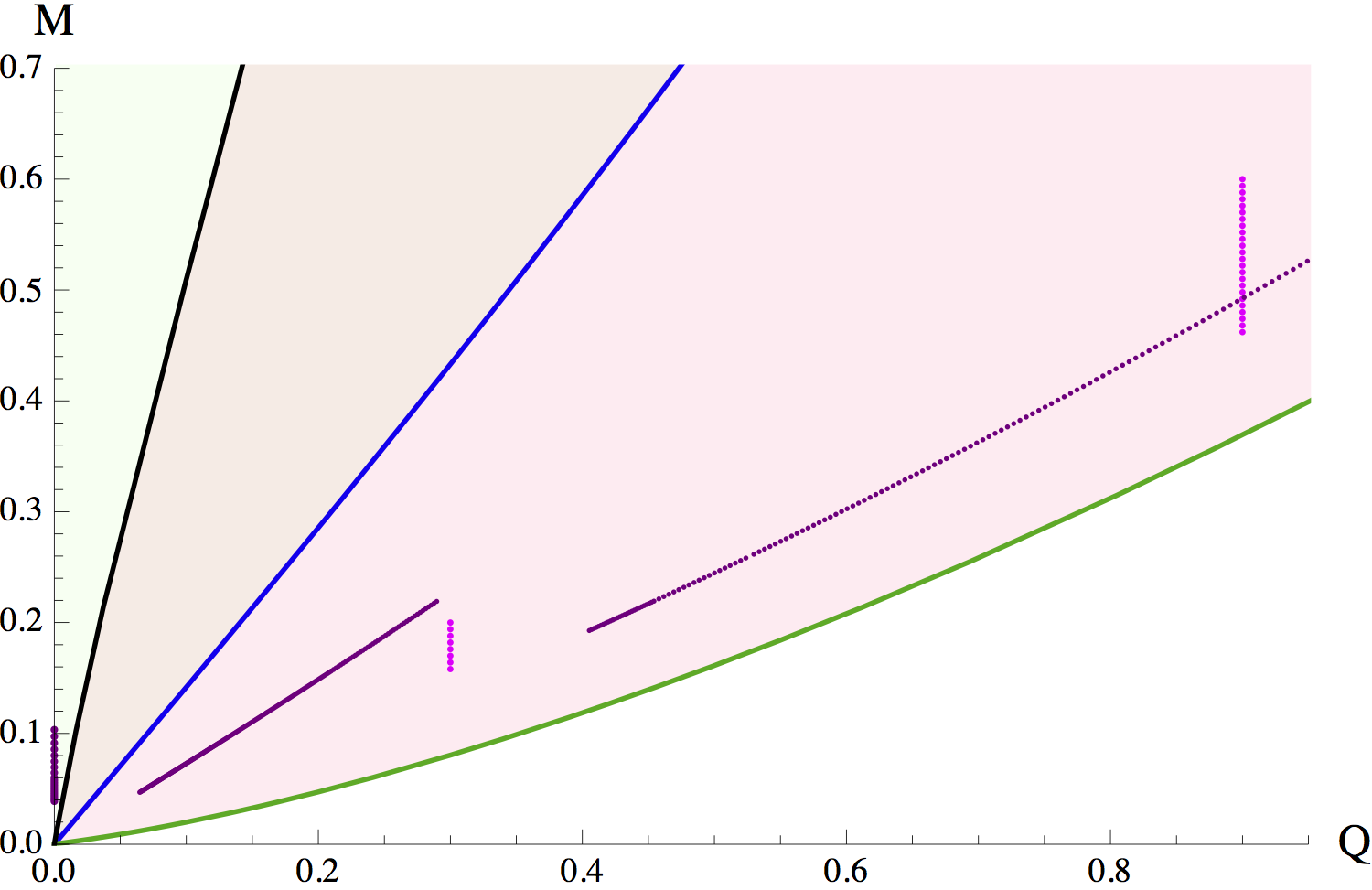}~~~~~\includegraphics[scale=0.65]{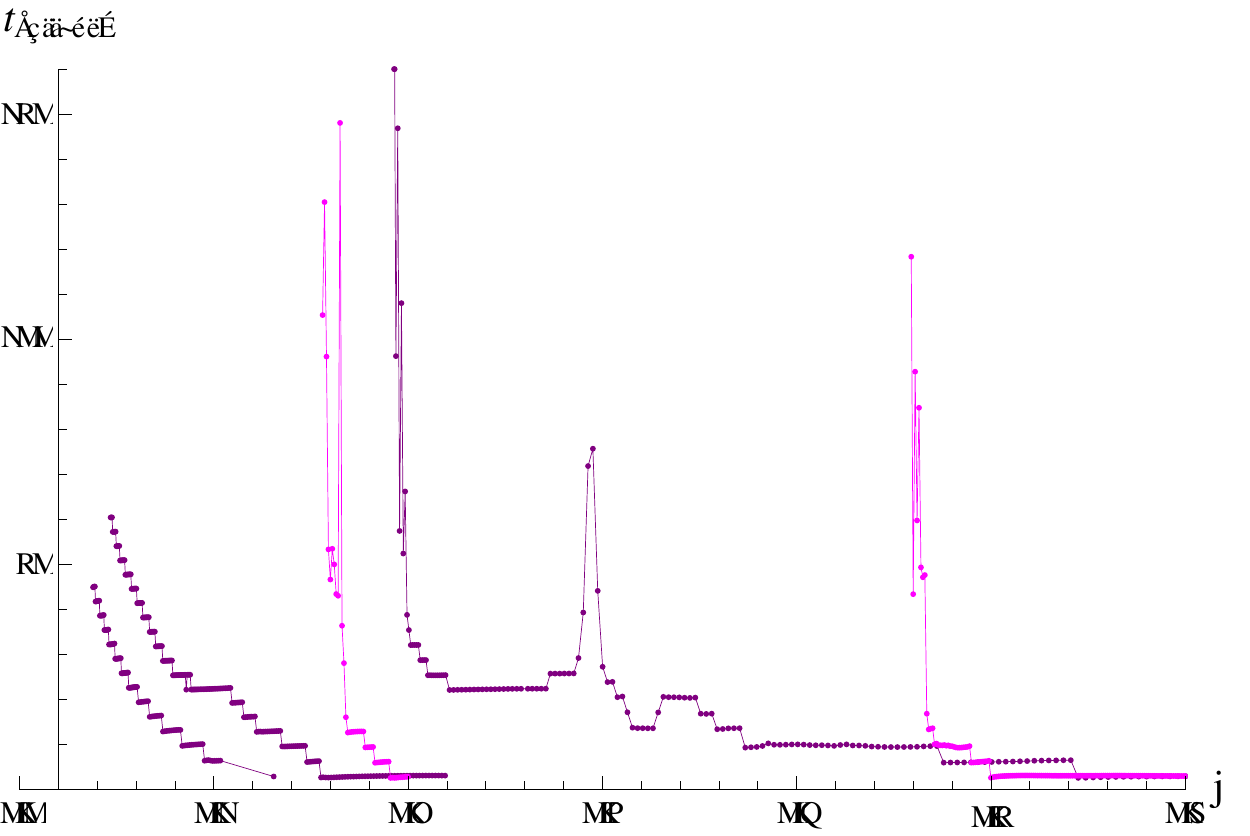}
\caption{\small  Collapses with $\sigma=0.2$ and $e=20$.
The dotted purple lines represent three series with  $\beta=0^\circ, 45^\circ$ and $63^\circ$. Vertical magenta  lines represent collapse series at constant $Q=0.3$ and $0.9$.  On the right plot the associated time curves are in direct correspondence from left to right.}
\label{fig:imecollapsesigma0d2e20}
\end{center}
\end{figure}

The $1/\epsilon^2$ scaling is apparent in the plots of figure \ref{fig:xscalingplot} for the maxima of the scalar curvature at the origin. The orange curves correspond to the orange dots in figure \ref{fig:imecollapsesigma0d1e20} and the same is true for the brown curves and dots.

After the works  in \cite{Dimitrakopoulos:2014ada, Craps:2014jwa, Buchel:2014xwa}, it has become clear that both focusing and defocusing dynamics (i.e. direct and inverse cascade)  seem to be in action and in a delicate equilibrium. For very sharp initial data, small $\sigma$, focusing wins.
In figure \ref{fig:imecollapsesigma0d2e20},  times for collapse with initial width $\sigma =0.2$ are plotted. For vanishing charge $\beta=0,\pi/2 \Rightarrow Q=0$, it scales indefinitely as expected as $1/\epsilon^2$ as for sharper pulses. However, as soon as some charge is added, we start seeing a divergence at finite values of $t_{collapse}$. This is exactly the same effect encountered in
\cite{Buchel:2013uba} but here it appears for smaller values of $\sigma$ than in that case.

\section{Soliton stability}

\noindent In recent years, configurations analogous to the solitons considered in this paper, known as  boson stars, have occupied a prominent role in the study of the AdS nonlinear stability problem. A boson star is a stationary complex scalar field configuration with non zero charge $Q$ that backreacts non trivially on the metric. However, in this case, the bulk U(1) symmetry is global and, therefore, boson stars carry no gauge field $A$. These solutions provide extended configurations that, once perturbed, help to shed light on how relevant an exactly resonant spectrum of linearized scalar field perturbations is for a weakly turbulent instability to be present on the system. This issue was raised in \cite{Dias:2012tq}, where the authors showed that just an asymptotically resonant spectrum is not enough to trigger a turbulent cascade along the lines of \cite{Bizon:2011gg}. Later, it was recognized that boson stars are not endowed with an exactly resonant spectrum and, furthermore, they were shown to be nonlinearly stable \cite{Buchel:2013uba} \cite{Maliborski:2014rma}. It remains to be seen if the correlation between these two facts survives the present situation. We expect a similar mechanism to be here behind the behaviour of the $\sigma = 0.2$ vertical lines in figure \ref{fig:imecollapsesigma0d2e20} which approach the soliton (green) line at constant charge $Q$. One of the main aims of this section is to provide evidence for this.

\subsection{Linear stability properties}

\begin{figure}[h!]
\begin{center}
\includegraphics[width=16cm]{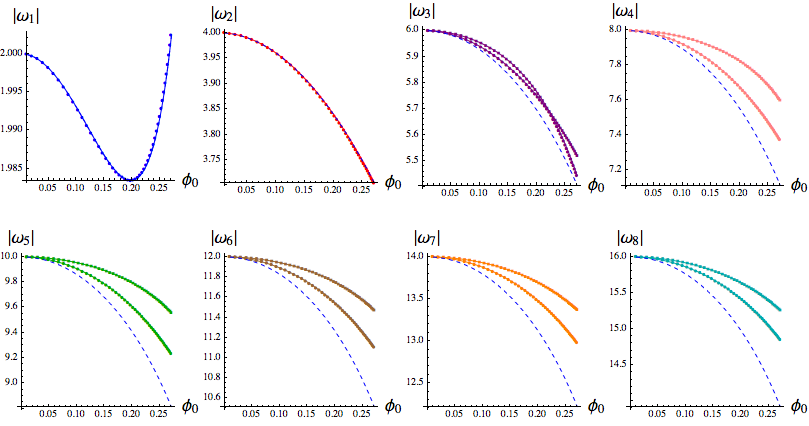}
\end{center}
\caption{\label{spectrum_dispersive_e=5} {\small Soliton scalar eigenfrequencies for $e = 5$.}}
\end{figure}

\begin{figure}[h!]
\begin{center}
\includegraphics[width=16cm]{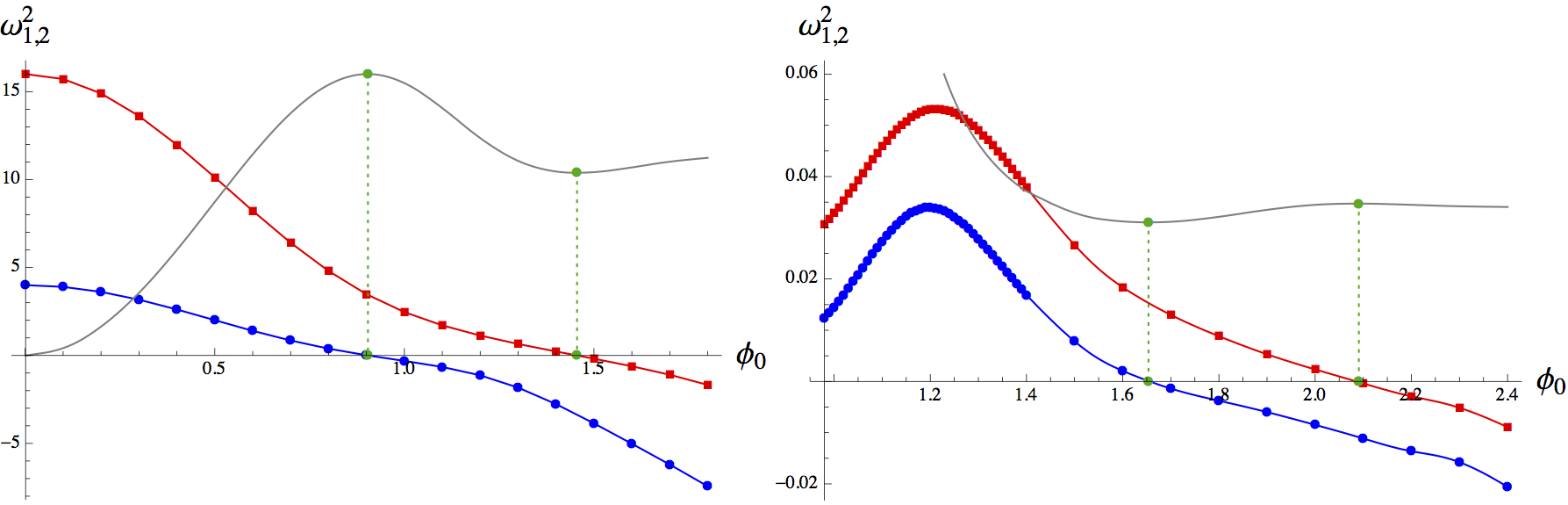}
\end{center}
\caption{\label{w_vs_phi0} {\small Left: $\omega_1^2$ (blue) and $\omega_2^2$ (red) versus $\phi_0$ for the vacuum connected soliton branch at $e = 2$. In grey we show the rescaled charge curve $Q(\phi_0)$. Right: $\omega_1^2$ (blue) and $\omega_2^2$ (red) versus $\phi_0$ for the vacuum disconnected soliton branch at $e = 2$. In grey we show the rescaled charge curve $Q(\phi_0)$. Green vertical lines correspond to the position of the charge curve extrema.}}
\end{figure}

\begin{figure}[h!]
\begin{center}
\includegraphics[width=16cm]{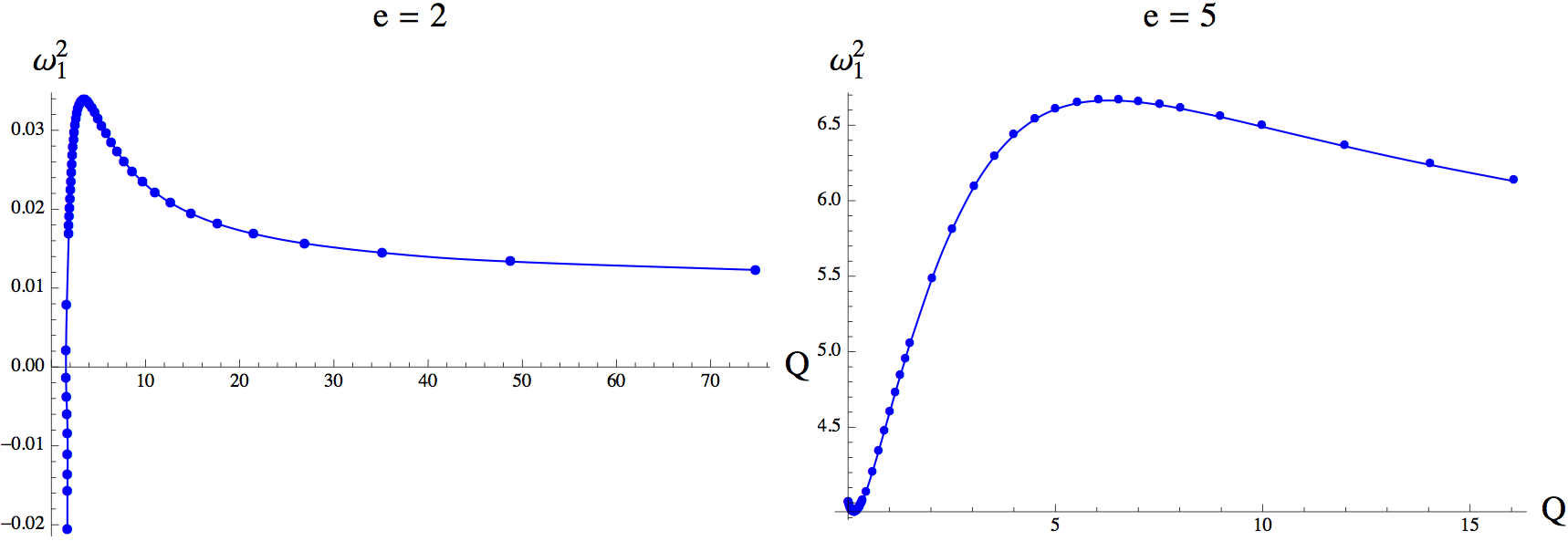}
\end{center}
\caption{\label{w_vs_Q} {\small Left: $\omega_1^2$ (blue) versus $Q$ for the vacuum connected soliton branch at $e = 2$. Right: $\omega_1^2$ (blue) versus $Q$ for the vacuum disconnected soliton branch at $e = 5$.}}
\end{figure}

\noindent   One of the aims of this subsection is to show  that whenever the soliton mass $M(\phi_0)$ attains an extremum the solutions become linearly unstable. In order to check this explicitly, in we are going to study linearized radial perturbations of the solitonic solutions with an harmonic time dependence of the form $\cos{ \omega t}$. Before delving into the details it is useful to notice that, as we are considering a time-reversal invariant and, therefore, nondissipative problem, $\omega^2$  is going to be purely real. In this way, an exponentially growing mode that signals an instability appears whenever $\omega^2 < 0$. We start by fixing our perturbations to be of the form\footnote{The strategy adopted in this paper is an adaptation of the method employed in \cite{Buchel:2013uba}\cite{Astefanesei:2003qy} to study boson star stability.}
\begin{eqnarray}
&&\delta(t, x) = \delta_s(x) + \epsilon \delta_1(t,x),  \label{d1def} \\
&&f(t, x) = f_s(x)(1 + \epsilon f_1(t,x)),  \label{f1def}  \\
&&A(t,x) = A_s(x) + \epsilon A_1(t,x),  \label{A1def}  \\
&&\phi(t,x) = \phi_s(x) + \epsilon (\phi_1(t,x) + i \phi_s(x) \partial_t \phi_2(t,x)), \label{phi1def}
\end{eqnarray}
with real $\phi_1, \phi_2$.  Since, due to  spherical symmetry, the metric carries no degrees of freedom in our setup, the perturbations defined by \eqref{d1def}-\eqref{phi1def} are not independent. In fact, the reason for having chosen this particular form for the scalar field perturbation is that it allows to solve  for $\delta_1$ and $f_1 $ in terms of $\phi_1, \phi_2$ and $A_1$, by making use of the momentum and Maxwell constraints \eqref{momentumconstraint},\eqref{cdotcons} linearized in $\epsilon$. Specifically, we get that
\begin{eqnarray}
&&\delta_1(t,x) = - \frac{\partial_x A_1(t,x)}{A_s'(x)} - \frac{2 e f_s(x) e^{-2 \delta_s(x)} \phi_s(x) \partial_x \phi_2(t,x)}{\cos(x)^2 A_s'(x)} + C_\delta(x), \label{d1redux} \\
&&f_1(t,x) = - \sin(2x) \left(\phi_s'(x) \phi_1(t,x) - e A_s(x) \phi_s(x)^2 \partial_x \phi_2(t,x) \right) + C_f(x), \label{f1redux}
\end{eqnarray}
where $C_\delta, C_f$ are integrating functions that must be fixed by the correct choice of boundary conditions. For harmonic perturbations, set
\begin{equation}
(\delta_1(t,x), f_1(t,x), A_1(t,x), \phi_1(t,x), \phi_2(t,x)) = (\hat{\delta}_1(x), \hat{f}_1(x), \hat{A}_1(x), \hat{\phi}_1(x), \hat{\phi}_2(x)) \cos {\omega t} \label{harmonicdef}
\end{equation}
which forces $C_\delta, C_f$ to be zero. Then, we can obtain the equations of motion for $\hat{\phi}_2$ and $\hat{A}_1$ by linearizing in $\epsilon$ the equations for $\delta$ and $A$, \eqref{Einstein} and \eqref{ceq} (the linearized equation for $f$ is not independent). The remaining equation for $\hat{\phi}_1$ comes from equation \eqref{eqmovPi}  for  $\phi$, linearized in $\epsilon$, after making use of both the $\hat{\phi}_2$ and $\hat{A}_1$ equations. The final expression of the equations of motion for the perturbations is not particularly illuminating. Defining
\begin{equation}
\vec{Z} =  \left(\hat{\phi}_1, \hat{\phi}_2, \hat{A}_1 \right)^T,  \label{Zdef}
\end{equation}
the perturbation equations are of the form
\begin{equation}
\vec{Z}''(x) + M^1(x) \vec{Z}'(x) + \left(M^{0}_a(x) + \omega^2 M^{0}_b(x) \right) \vec{Z}(x) = 0, \label{pertdef}
\end{equation}
where $M^1, M^0_{a,b}$ are matrix-valued functions that depend exclusively on the background solution. In order to solve \eqref{pertdef}, we have to choose appropriate boundary conditions for $\vec{Z}$. At $x = 0$ we demand regularity. As for the background soliton, this forces $\vec{Z}$ to be even at $x = 0$. At  $x = \pi/2$, boundary conditions are
\begin{eqnarray}
&&\hat{\phi}_1(x) = O(\rho^3),  \label{bc1} \\
&&\hat{\phi}_2 = \hat{\phi}_{2,0} + O(\rho^2), \label{bc2} \\
&&\hat{A}_1 = O(\rho^2). \label{bc3}
\end{eqnarray}
The first two conditions come from imposing normalizability on the scalar field perturbation  \eqref{phi1def}. The last condition demands a more thoughtful explanation. Let us consider the most general near boundary expansion for $\hat{A}_1$,
\begin{equation}
\hat{A}_1(x) = \hat{A}_{1,0} + \hat{A}_{1,1} \rho + O(\rho^2).
\end{equation}
In this case, it can be shown that we  have $\hat{\delta}_1(\pi/2) \propto \hat{A}_{1,1}$ and, in consequence, if we want to maintain our gauge choice for the time coordinate, we must set $\hat{A}_{1,1}=0$. This is tantamount to demanding that the frequency $\omega$ is the one measured by a boundary observer.\footnote{Under a shift $\delta(x) \rightarrow \delta(x) + c$, $\omega$ changes as $\omega \rightarrow \omega e^{-c}$ so as to maintain the $\omega t$ phase of the perturbation invariant.} On the other hand, nothing prevents us from allowing that $\hat{A}_{1,0} \neq 0$ i.e.,  perturbations that don't keep fixed the soliton chemical potential. However, examining the explicit form of equation \eqref{pertdef} we discover that any solution is invariant under the change $\hat{\phi}_1 \rightarrow \hat{\phi}_1$, $\hat{\phi}_2 \rightarrow \hat{\phi}_2 + \alpha$ and $\hat{A}_1 \rightarrow \hat{A}_1 + \beta$, provided that $\alpha \omega^2 + e \beta = 0$. Therefore, we can employ this residual symmetry\footnote{This residual symmetry stems from the fact that equation \eqref{pertdef} is a linear ODE that admits an algebraic solution of the form $\vec{Z} = \left(0, \alpha, \beta \right)^T$ with the aforementioned coefficient choice. This algebraic solution, on the other hand, is nothing but the action of a linearized gauge transformation on the trivial $\vec{Z} = 0$  solution.} to set $\hat{A}_{1,0} = 0$ with no loss of generality, fixing it completely along the way.\footnote{We have checked explicitly that the numerical results presented further on are independent of the particular way this symmetry is fixed. They are also equivalent to the ones obtained when this symmetry is left unbroken.}
\\\\
\noindent Before discussing how equation \eqref{pertdef} was solved numerically, let us make a last general comment. The boundary conditions \eqref{bc1}-\eqref{bc3} and the relations \eqref{d1redux}-\eqref{f1redux} imply that the perturbations here considered don't change the charge and the mass of the soliton at linear order. This observation allows for a better understanding of the relation between the soliton linear stability properties and the fact that the mass curve, $M(\phi_0)$, encounters an extremum at $\phi_0 = \phi_{0,c}$.\footnote{See also the related discussion in \cite{Bhattacharyya:2010yg}.} First, let us mention that, whenever $M'(\phi_{0,c})=0$, we also find that $Q'(\phi_{0,c})=0$. Therefore, around $\phi_{0,c}$, two infinitesimally close solitons, parameterized respectively by $\phi_{0,c}$ and $\phi_{0,c} + \Delta \phi$, have the same mass and charge, up to $O(\Delta \phi^2)$ corrections. This implies that there must be a time-independent linear radial perturbation that connects these static configurations and, in consequence,  equation \eqref{pertdef} admits a solution with $\omega_1^2(\phi_{0,c}) = 0$, i.e. a zero mode in the soliton spectrum. For  $\omega^2_1(\phi_0)$  at least a $C^2$ function around $\phi_{0,c}$, we find
\begin{equation}
\omega_1^2(\phi_0) = \omega_1^2(\phi_{0,c}) + \partial_{\phi_0} \omega_1^2(\phi_{0,c}) (\phi_0 - \phi_{0,c}) + ... = \partial_{\phi_0} \omega_1^2(\phi_{0,c}) (\phi_0 - \phi_{0,c}) + ...
\end{equation}
which becomes negative one side of the mass curve extremum, signalling an instability. On the remaining part of this section we are going to proceed by solving \eqref{pertdef} numerically, confirming this expectation.
For this task we employed  Tchebychev pseudospectral collocation method. Inserting expansions
\begin{eqnarray}
&&\hat{\phi}_1(x) = \cos(x)^3 \sum_{k=0}^{N-1} c_{1,k} T_k(1- 4/\pi x),  \label{ansatz1} \\
&&\hat{\phi}_2(x) = \sum_{k=0}^{N-1} c_{2,k} T_k(1- 4/\pi x),  \label{ansatz2} \\
&&\hat{A}(x) = \cos(x)^2 \sum_{k=0}^{N-1} c_{3,k} T_k(1- 4/\pi x), \label{ansatz3}
\end{eqnarray}
 into equation \eqref{pertdef}  and evaluating on a collocation grid $\{t_k, k = 1 ... N \}$
\begin{equation}
t_k = \frac{\pi}{4} \left(1 - \cos\left(\left(k - \frac{1}{2}\right) \frac{\pi}{N} \right) \right),  \label{griddef}
\end{equation}
 an algebraic generalized eigenvalue problem is to be solved which gives the numerical values of the first soliton normal modes.\footnote{We discard the values that don't converge when $N$ is increased. A convergence test for the pseudospectral code is provided in appendix D.}
\\\\
\noindent  As for the results, first, the spectrum thus found is not resonant. In the spirit of \cite{Dias:2012tq}, this should entail the absence of a turbulent cascade  in the fully nonlinear regime. In figure \ref{spectrum_dispersive_e=5}, we plot the first eight normal frequencies for the soliton branch at coupling $e = 5$. They consistently reduce to their global AdS values as $\phi_0 \rightarrow 0$. A remarkable feature is the mode splitting that occurs for $k \geq 3$.\footnote{We remind the reader that a mode splitting was previously found in the perturbative computation of \cite{Maliborski:2014rma}.} This is common to every perturbative soliton branch we have analyzed. If the spectrum was exactly resonant, we would have that, for $k\geq2$,
\begin{equation}
|\omega_{k+1}| = |\omega_1| + k (|\omega_2| - |\omega_1|). \label{eqresonant}
\end{equation}
We plot the right hand side of \eqref{eqresonant} for each $k$ in figure \ref{spectrum_dispersive_e=5} (blue-dashed). It is clearly seen that the equality is not satisfied away from $\phi_0 = 0$. A similar exercise can be performed only between the lower or upper splitted eigenfrequencies, choosing as reference the difference between any two consecutive ones, with identical conclusion.
\\\\
\noindent Let us consider the intermediate region $3/2\leq e^2\leq 9/2$. In figure \ref{w_vs_phi0} (left), we plot $\omega_1^2(\phi_0), \omega_2^2(\phi_0)$ for the vacuum connected soliton branch at $e=2$, together with the rescaled $Q(\phi_0)$ curve. We clearly see that a zero mode develops precisely at the point where $Q(\phi_0)$ reaches its first maximum and that, past this point, the solutions become linearly unstable.\footnote{With a resolution of $\delta \phi_0 = 10^{-3}$, we have determined that the maximum lies at $\phi_{0,c} = 0.904$. Our pseudospectral code produces the values $\omega_1^2(\phi_0=0.903) = 0.0031$ and $\omega_1^2(\phi_0=0.905) = -0.0037$, in perfect agreement with expectations.} The same phenomenon can be clearly appreciated on the vacuum disconnected soliton branch after $Q(\phi_0)$ attains its first minimum (figure \ref{w_vs_phi0}, right).  The general pattern we find is that, when $Q(\phi_0)$ hits a new extremum, a new normal mode crosses zero.\footnote{This is a non trivial result obtained from the computation, since apparently nothing prevents the other option, where some normal mode oscillates up and down, crossing zero at every extremum of the charge curve.} \\
\\
\noindent As an additional comment, note that, regarding the vacuum disconnected branch, and in the $Q\gg 1$ regime, $\omega_1^2(Q)$ is a decreasing function of $Q$ that stays finite in the $Q \rightarrow \infty$ limit (figure \ref{w_vs_Q}, left). Instead, in order for the phase of the harmonic perturbation $\omega t$ to remain finite in the blow up limit \eqref{blow_up_map}, the frequency should scale up as  $\omega \sim \mu \sim Q^\frac{1}{2}$. We conclude that harmonic linear perturbations die off when the blow up limit is taken. This is consistent with the fact that the soliton branch maps onto a $T = 0$ hairy black brane, for which linearized perturbations correspond to quasinormal rather than normal modes. The discussion goes through in parallel to the regime $e > e_{sr}$ (see figure \ref{w_vs_Q} (right) for  $\omega_1^2(Q)$ at $e = 5$).
\\
\\
\noindent After having discussed the linear stability properties of the solitons, in the next section we move on to the study of their nonlinear stability.

\subsection{Nonlinear stability properties}

\noindent We consider now the effect that a localized scalar field perturbation has on the soliton. We will stick to the same family of initial conditions that were used to perturb the $AdS_4$ vacuum \eqref{gauss1}\eqref{gauss2} but
in a purely real setup. Concretely, our initial condition will be  $\phi_s + \phi$ with  $\phi(0,x)=0$ and
\begin{equation}
\Pi_1(0, x) =\epsilon \frac{2}{\pi} \exp\left(- \frac{4 \cot^2(x)}{\pi \sigma^2} \right) \cos^3(x)~~,~~~  \Pi_2(0,x)=0\, . \label{pertid}
\end{equation}
As this  configuration has zero charge (see \eqref{qfromat}), the family of perturbed solitons that we use to start with spans a vertical line in the $(Q, M)$ plane above the unperturbed soliton solution $\phi_s$,  like the magenta vertical
sets in figure \ref{fig:imecollapsesigma0d2e20} (left). The difference now is that this set of initial conditions explores down to the bottom green line as $\epsilon\to 0$.

\begin{figure}[h!]
\begin{center}
\includegraphics[width=16cm]{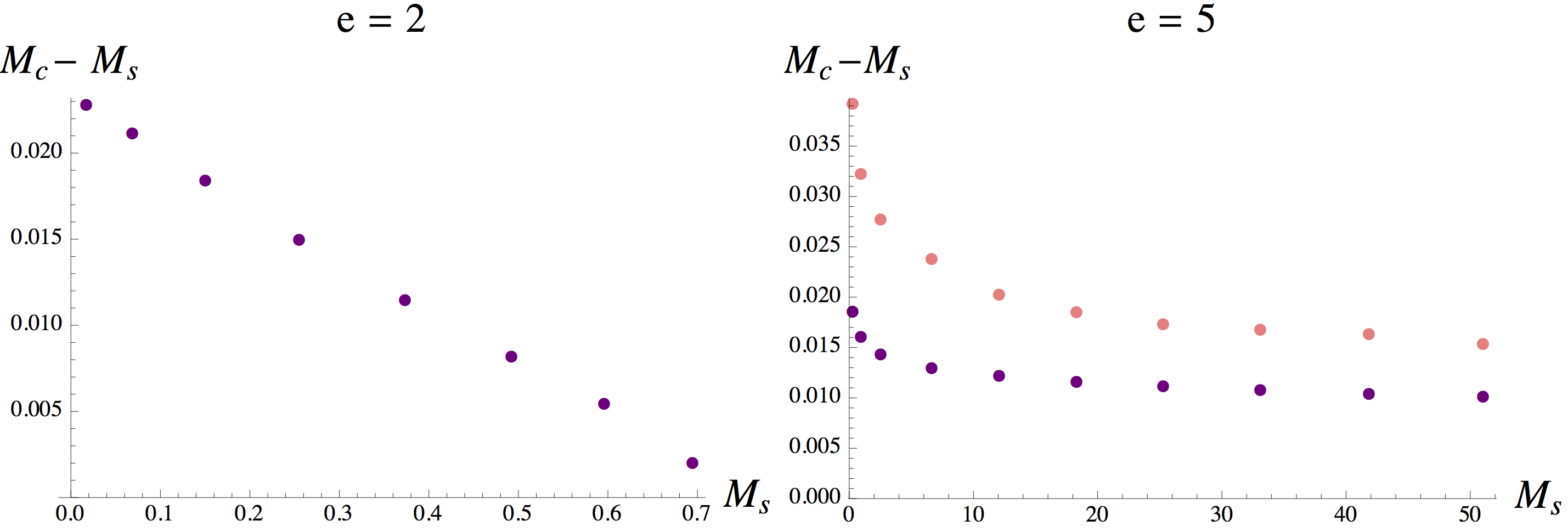}
\end{center}
\caption{\label{Fig3} {\small Left: limit of the prompt collapse region for the vacuum connected $e = 2$ soliton branch for $\sigma = 0.05$ (purple). Right: limit of the prompt collapse region for the $e = 5$ soliton branch for $\sigma = 0.1$ (magenta) and $\sigma = 0.05$ (purple).}}
\end{figure}

Despite the fact that placing a perturbation like \eqref{pertid} on top  of  the AdS vacuum or on top of a soliton leads to very different initial conditions, the phenomenology we discover is remarkably similar. Namely, the magenta lines found in figure \ref{fig:imecollapsesigma0d2e20} (right) are
qualitatively reproduced  here. Indeed, for high enough $\epsilon$ prompt collapse is observed. Below some threshold mass $M_c$ we have a delayed collapse and a number of  oscillations are completed before the system finally undergoes gravitational collapse. This number, and with it the final time for collapse, diverges rapidly at some value  of the mass above the soliton curve. Hence we don't see any trace of a nonlinear instability  corner centered at the soliton solution (instead of the AdS vacuum).
This is presumably again a symptom of the nonresonant character of the spectrum of soliton perturbations.

\noindent
The possible survival  of this oscillating region in the blow up limit is a relevant question and, indeed, was one of the main motivations that started the present work; the answer is negative.  More precisely, we are interested in establishing whether  its width $M_c - M_s$ has a finite size relative to $M_s$
\begin{equation}
M_c - M_s \sim M_s \sim Q_s^{2/3}.
\end{equation}
In figure \ref{Fig3} (right) we plot the location of the mass threshold $M_c - M_s$ for prompt collapse for the $e=5$  family of solitons of mass $M_s$. We consider perturbations of the form \eqref{pertid} with $\sigma = 0.1$ and  $\sigma = 0.05$. We see that, despite having finite width for large mass, the  region of oscillations does not have the appropriate scaling to survive in the blow up limit. In this case, for $M_s \gg 1$, entering the oscillating regime requires to fine-tune the initial perturbation \eqref{pertid} in such a way that its relative contribution to the system energy goes to zero.
\noindent The behaviour just described is fairly natural given that, in the blow up limit, the soliton branch we are perturbing reduces to an extremal hairy black brane with an AdS$_4$ near-horizon geometry \cite{Horowitz:2009ij}. It is not unreasonable that, above this background, any perturbation localized in the near-boundary region, regardless of its amplitude, leads to direct collapse to a $T \neq 0$ black brane, in the same vein as it happens for zero charge  \cite{Wu:2012rib}. It remains to be seen if this is also the pattern in other matter models. Our educated guess is that, for any theory that displays an unbounded soliton branch, the blown up extension of the oscillatory regime is finite whenever the theory can support a gapped spectrum of scalar fluctuations in the planar limit. Theories of this kind are not unknown. Consider, for instance, the Improved Holographic QCD models which have recently been analyzed at the dynamical level in \cite{Ishii:2015gia} \cite{Ishii:2016uxe}. There, the planar geometry dual to the field theory ground state is sourced by a nontrivial scalar field profile that generates a naked singularity in the infrared.\footnote{This, and what follows in the main text, is true at least for some specific classes of scalar potentials.} When considered in the fully nonlinear regime, weak perturbations localized near the boundary may be noncollapsing and forever oscillating, since the singularity would ``repel" them from the infrared, so as they never reach their Schwarzschild radius. This behavior has been explicitly seen in more crude models of gapped field theories in planar AdS, such as a scalar field in a hard wall geometry \cite{Craps:2014eba} or the AdS-soliton \cite{Craps:2015upq}. It would be interesting to classify, in generic terms, which Einstein-Maxwell-scalar theories support a soliton branch with a gapped planar limit and check if this is correlated with a nonvanishing scaling of the oscillatory regime width.\\\\
\noindent We have also determined the boundary of the prompt collapse region for the vacuum connected soliton branch at coupling $e = 2$, upon scalar fluctuations of the form \eqref{pertid} with $\sigma = 0.05$ (figure \ref{Fig3}, left). In accordance with the fact that soliton solutions become linearly unstable when the mass curve reaches its maximum, here we find that the width of the oscillation region shrinks to zero in a linear fashion.

\section{Conclusions}

The original motivation of this paper was to examine the status of the nonlinear instability problem
in  the Einstein-Maxwell-scalar theory in $d+1=4$ dimensions.  The work demanded first a thorough unraveling of the
landscape of static solutions which showed almost the same features as in one dimension higher \cite{Basu:2010uz}\cite{Dias:2011tj}.
The issue of stability can be examined at different levels. Studying the thermodynamical linear stability  demands
 selecting first a certain ensemble. Working in the grand canonical ensemble we have pinned down the relevant phase transitions.

In the second part of the paper we have studied the nonlinear stability in the microcanonical ensemble. First of all, we have considered perturbations of pure AdS$_4$ with families of initial conditions of different charges, generalizing the uncharged ones considered in  \cite{Bizon:2011gg}.
These initial conditions have an amplitude $\epsilon$ and a width $\sigma$.
We have found that for thin initial pulses $\sigma =0.1$, the same type of instability sets in at times of order $\epsilon^{-2}$ .
This fact   confirms the expectation that places the origin of the mechanism in the fully resonant character of the linear spectrum,
something that is also true in this case.
Wider pulses, $\sigma\geq 0.2$, exhibit a divergence in the time for collapse below some critical amplitude.  This is also in parallel with the effect detected for wide radial  perturbations of  AdS$_4$ in \cite{Buchel:2013uba}.
The role of the oscillons in that situation is probably taken up by the solitons in this setup.

To confirm this picture,
 we have examined the linear and nonlinear stability of fluctuations placed on top of the soliton solutions. The spectrum of linear normal modes
is not resonant. The linear stability, signalled by an imaginary eigenfrequency,  is seen to appear, as expected, coinciding with extrema of the mass curve.
For higher amplitude perturbations we need to resort to a full fledged simulation of the evolution of the system.
In general, linear stability extends to nonlinear stability up to some threshold for the amplitude of the perturbation.
Beyond that, the soliton is destroyed and collapses to a hairy black hole. The protection region where oscillations do not decay does not scale
properly in the limit of large mass and charge to survive the blow up planar limit. This seems to point to the necessity of having a mass gap, and not just a mass scale,  to find such oscillatory behaviours.

While this paper was  completed in its last part, namely sections 4 and 5, reference \cite{Basu:2016mol} appeared in the arXive. It overlaps substantially with the content of our section 3, and we find agreement with their results whenever they can be compared (their scalar field has negative mass).

\section{Acknowledgements}

We want to thank Oscar Dias, Pau Figueras, Esperanza L\'opez and Gonzalo Torroba for useful conversations and Angel Paredes for a carefully reading of the manuscript. R.A. is supported by CONICET and Universidad Nacional de Cuyo. The work of J.M. is supported in part by the spanish grants FPA2011-22594 and FIS2014-61984-EXP, by Xunta de Galicia (GRC2013- 024), by the Consolider-CPAN (CSD2007-00042),   and by FEDER. A.S. is supported by the European Research Council grant HotLHC ERC-2011-StG-279579 and by Xunta de Galicia (Conselleria de Educaci\'on). The work of R.A. and J.M. was partially supported by EPLANET program. This research has benefited from the use computational resources/services provided by the Galician Supercomputing Centre (CESGA).

\appendix
\vskip 1cm
\setcounter{equation}{0}
%\medskip

\section{Ansatz and equations of motion}\label{apansatz}

Let us give the equations for a situation which is slightly more general than the one considered in the main text, and allow for a scalar potential
\be
S= \frac{1}{2\kappa^2} \int d^{d+1}x \sqrt{-g}\left( R - 2\Lambda\right)   -\int d^{d+1}x\sqrt{-g}\left(   D_\mu\phi \bar D^\mu\phi^*  + V(\phi)\right) -\frac{1}{4}\int d^{d+1} x \sqrt{-g} F_{\mu\nu} F^{\mu\nu}
\ee
with $\kappa^2 = 8\pi G$,  $\Lambda = -d(d-1)/2l^2$ for Anti de Sitter. Also, the scalar field is complex and we have $D_\mu \phi \equiv (\pd_\mu - i e A_\mu)\phi$.
The equations of motion are
 \beqa
 R_{\mu\nu} - \frac{1}{2} g_{\mu\nu} R +\Lambda g_{\mu\nu} &=& \kappa^2\left(T_{\mu\nu}^{(\phi)} + T_{\mu\nu}^{(A)}, \right)\label{eqeins}\\
\frac{1}{\sqrt{-g}} D_\mu \left( \sqrt{-g} g^{\mu\nu} D_\nu\phi\right) &=&  \frac{\pd V(\phi,\phi^*)}{\pd\phi^*}, \label{eqphi}\\
\frac{1}{\sqrt{-g}}\pd_\mu (\sqrt{-g} F^{\mu\nu} ) &=&  J^\nu, \label{MaxwEq}
 \eeqa
 where the energy-momentum and charge currents are
 \beqa
 T_{\mu\nu}^{(\phi)} &=& \left( \bar D_\mu\phi^* D_\nu\phi+ \mu \leftrightarrow \nu\right) - g_{\mu\nu} \left( |D\phi|^2
+ V(\phi)\right), \\
 T_{\mu\nu}^{(A)} &=& F_{\mu\alpha} F_{\nu}{^\alpha} - \frac{1}{4} g_{\mu\nu} F^2,  \\
 J^\nu &=&  i e \left( \phi^* D_\mu \phi - \phi (\bar D_\mu\phi^*) \right) g^{\mu\nu}.
 \eeqa

By employing the ansatz described in section 2, the Klein-Gordon equation can be casted into the first order form
\beqa
\dot\Phi &=& \left(f e^{-\delta}  \Pi + ie A_t \phi\right)',  \label{eqmovPhi} \\
\dot\Pi &=&   \displaystyle \frac{1}{\tan^{d-1} x}\left(\tan^{d-1} x f e^{-\delta} \Phi\right)' + ie A_t \Pi- \frac{l^2}{\cos^2x} e^{-\delta} \partial_{\phi_c} V(\phi ), \label{KG} \label{eqmovPi}
\eeqa
while, from the Einstein equations, we obtain,
\beqa
f' &=&\displaystyle \frac{d-2 + 2 \sin^2 x}{\sin x \cos x} (1-f) -\frac{2\kappa^2}{d-1}\sin x \cos x \, f\,  \left( |\Phi|^2 + |\Pi |^2   \right)
\label{Einstein1}\\
&&- \displaystyle{\frac{\kappa}{(d-1)l^2}} e^{2\delta} \cos^3 x \sin x \, A_t'(t,x)^2 - \frac{2\kappa^2 l^2}{d-1} \tan x \, V(|\phi |), \nonumber \\
\delta' &=&-\displaystyle  \frac{2\kappa}{d-1} \sin x \cos x \left(  |\Phi|^2 + |\Pi |^2 \right). \label{Einstein2}
\eeqa

There is one additional equation coming from the $(t,x)$ component of Einstein equations that yields the {\em momentum  constraint}
\be
\dot f = -\frac{4\kappa}{d-1} \sin x \cos x f^2 e^{-\delta} \, {\rm Re}(\Phi\Pi_c). \label{momentumconstraint}
\ee

Concerning  Maxwell's equations,  define $C = A_t'(t,x)$, then from \eqref{MaxwEq} we can derive the two following equations
\beqa
{\left(e^{\delta} C\right)}' &=& 2e\frac{ l^2}{\cos^2 x} {\rm Im}(\phi\Pi_c) - e^{\delta} C ((d-3)\tan x  +(d-1)\cot x)
 \\
\dot {\left(e^{\delta} C\right)} &=&2e \frac{ l^2 e^{-\delta} f}{\cos^2 x} {\rm Im}(\phi\Phi_c)
\label{cdotcons}
\eeqa
which can be easily shown to be compatible. The first one can be recasted as follows
\be
\left( e^{\delta} C   \sin^{d-1} x \cos^{3-d} x \right)' = 2e l^2 \tan^{d-1} x\,  {\rm Im}(\phi\Pi_c)
\label{ceq}
\ee
and the second   is the {\em  Maxwell constraint}.
 With the  condition that $A'_t$ be bounded at the origin we can integrate \eqref{ceq} to find
\be
A'_t =  2 e l^2  e^{-\delta} \sin^{1-d} x \cos^{d-3} x  \,  \displaystyle \int_{x_0}^x \tan^{d-1} x\,  {\rm Im}(\phi \Pi_c)
\label{intat}
\ee

The electromagnetic current is given by the following expression
$$
J^{\mu} (t,x) = \frac{2e\cos^2 x  }{l^2}   (- e^\delta \,{\rm Im}  (\phi\Pi_c ), f   \,{\rm Im}  (\phi\,\Phi_c ),0,0)
$$
and the charge density leads to the definition of the conserved {\em physical} charge
\beqa
{\cal Q} &=&\int_{S_{d-1}} d\Omega_{S_{d-1}}\int_0^{\pi/2}dx  \sqrt{-g} J^0(x) \\
&=&
-2e  l^{d-1} V_{S_{d-1}} \int_0^{\pi/2} dy  \tan^{d-1}y  \,    {\rm Im} (\phi\Pi_c )
\eeqa

Take \eqref{intat} and notice that the following expression holds
\be
\frac{(-1)^{d-3}}{(d-3)!} A_t^{(d-2)} (\pi/2)= 2 e l^2  e^{-\delta(\pi/2)} \int_0^{\pi/2} \tan^{d-1} y \, {\rm Im}(\phi \Pi_c) = \frac{-{\cal Q} e^{-\delta(\pi/2)}}{l^{d-3} V_{S_{d-1}}}
\ee
or, after dividing by $(2-d)$ and multiplying by $e^{\delta(\pi/2)}$, because ${\cal Q}$ has to be invariant under $\delta \to \delta + c$,
\be
e^{\delta(\pi/2)}\frac{(-1)^{d-2}}{(d-2)!} A_t^{(d-2)} (\pi/2)= \frac{  {\cal Q}}{(d-2)l^{d-3} V_{S_{d-1}}} \equiv Q.\label{eqQ}
\ee
In other words
\be
Q = -\frac{2e l^2}{d-2}  V_{S_{d-1}}  \int_0^{\pi/2} \tan^{d-1} y\,  {\rm Im}(\phi \Pi_c).  \label{qfromat}
\ee

The Reissner-Nordstr\"om black hole solution can be recovered by setting $\phi=\Phi=\Pi = \delta = 0$. Specializing to the $d=3$ case, the stationary solution to \eqref{Einstein} and \eqref{ceq} is
\beqa
f_{RN}(x)  &=& 1 - M \frac{\cos^3 x}{\sin x} + \kappa\frac{1}{2} Q^2\frac{\cos^{4} x}{\sin^{2} x}, \label{efern}\\
A_{t\, RN}(x) &=&  \mu  + Q \frac{\cos x}{\sin x}. \label{nb1}
\eeqa
This expression gives the mass and charge of the black hole as
\beqa
M &=& \frac{1}{6} \, f^{(3)}_{RN}(\pi/2) \label{MRN}\, ,\\
Q&=& -A_t'(\pi/2)e^{\delta(\pi/2)}\label{Q} \label{nb2} \, .
\eeqa
The charge is always proportional to the derivative of the gauge field on the boundary, i.e. the charge density, which agrees with \eqref{eqQ}.
More generally, the mass of a black hole in the time dependent background \eqref{line1} can be written
\beqa
M &=&(\sin x_h)(\sec x_h)^3 e^{\delta(\pi/2)-\delta(x_h)} ~+  \label{Mass}\\  \rule{0mm}{8mm}
&& +    \kappa \, e^{\delta(\frac{\pi}{2})} \int_{x_h}^{\pi/2}\left(\tan^{2}y  \left( \Phi^2 + \Pi ^2 + \frac{V(\phi)}{\cos^2 x} \right)  e^{-\delta(y)}
+ \frac{1}{2l^2}\sin^{2}y A_t'(t,y)^2e^{\delta(y)}\right)\nonumber
\eeqa

\section{Instabilities }

Take the RN metric
\be
ds^2 = -f(r) d  t^2 + \frac{dr^2}{f(r)}  + r^2 d \Omega^2_{d-1}
\ee
with
\beqa
f(r) &=& 1+ \frac{r^2}{l^2} -  M  \left(\frac{l}{r}\right)^{d-2} +\kappa^2\frac{d-2}{d-1} Q^2\left(\frac{l}{r}\right)^{2d-4},
\\
A_{t}(r) &=& \mu \left( 1- \frac{r_h^{d-2}}{r^{d-2}}\right) = \mu -\frac{Q}{r^{d-2}}. \label{mudeQ}
\eeqa
We will show that extremal RN black holes suffer from two potential instabilities.

\subsubsection{Tachyonic instability}

This is sourced by the lowest scalar mode close to the horizon. The near horizon limit is obtained by expanding
\be
r \to r_h (1 + \lambda/z)~~~, ~~~~t \to \left( \frac{f''(r_h)}{2} r_h\lambda  \right)^{-1} t  \label{NHLimit}
\ee
and taking the limit $\lambda\to 0$.  In the extremal case $f(r_h)=f'(r_h)=0$ and, therefore, the limit has the form of AdS$_2\times S^{d-1}$
$$
ds^2 \to \left(\frac{f''(r_h)}{2}\right)^{-1} \left( \frac{ -dt^2 + dz^2}{z^2}\right)  + r_h^2 d\Omega_{d-1}^2
$$
and also
$$
A_t \to   \frac{\lambda \mu (d-2)}{z}\, .
$$

One may take the near horizon limit in the equation of motion of the scalar field and extract the mass from the coefficient of $\phi$
\be
m^2 = -\frac{4 e^2(d-2)^2\mu^2}{f''(r_h)^2 r_h^2}\, .
\ee
We observe explicitly  that the mass gap is set by the chemical potential.
The scalar field will be unstable if its mass violates the BF bound of AdS$_2,$ hence if $m^2< -1/4$. This implies
$$
e>e_t = \frac{1}{4}\frac{f''(r_h) r_h}{(d-2) \mu}
$$
The values of $M$ and $Q$ at extremality are given by
$$
M_{ext} =2 r_h^{d-2}\left(1 +  \frac{ d-1}{d-2} r_h^2 \right)~~~~;~~~~Q_{ext} = \mu_{ext}r_h^{d-2} = \frac{r_h^{d-2}}{d-2}\left( \rule{0mm}{4mm}(r_h^2 d + d - 2)(d-1))    \right)^{1/2}
$$
Evaluating $f''(r_h)$ with these values results in the following expression
$$
e_t^2 = \frac{4 + d(d-4 + r_h^2(d-1))}{4 r_h^2(d-1)(d (r_h^2 + 1)-2)},
$$
which diverges in the limit $r_h\to 0$. The tachyonic instability is thus suppressed for infinitesimally small RN back holes. Instead, for large $r_h$, $e_t(r_h)$ admits the following expansion
\be
e_t^2 = \frac{d(d-1)}{4} + \frac{6-5 d + d^2}{4 r_h^2} + {\cal O}(r_h^{-4}) + . ..
\ee

In summary, for $e^2_t < 3/2$ in $d=3$ and for $e^2_t < 3$ in $d=4$,  there is no tachyonic instability. As this threshold is surpassed, large black holes are unstable first. Small black holes do not suffer from this instability.

\subsubsection{Superradiant instability}

If $e\mu >\Delta_0$, with $\Delta_0$ being the lowest eigenvalue of the scalar linearized equation, RN black holes becomes superradiant in the limit $r_h \rightarrow 0$, meaning that this mode scatters off with a reflection coefficient $|R| >1$. The final state should be a hairy black hole with greater entropy than the RN one, which dominates the micro-canonical ensemble. For a massless scalar, $\Delta_0 = d$. Substituting the extremal value
$$
\mu_{ext} = Q_{ext} r_h^{2-d} = \frac{1}{d-2}\left( \rule{0mm}{4mm}(r_h^2 d + d - 2)(d-1))    \right)^{1/2}
$$ gives, for $r_h \ll 1$,
\be
e^2_{sr} = \frac{d^2(d-2)}{d-1} \left( 1 -\frac{d}{2(d-2)}r_h^2 + {\cal O}(r^{3/2}) + ... \right),
\ee
which constitutes the relevant lower bound on the coupling, since $\mu < \mu_{ext}$. The precise superradiance threshold is
\begin{itemize}
\item
in $d=3$,
$$
e^2_{sr} = \frac{9}{2} -\frac{27 r_h^2}{2} + ...
$$
Hence all extremal RN black holes above $e^2=9/2$ are unstable.
\item
in $d=4$,
$$
e^2_{c} =  \frac{32}{3}  - \frac{64 r_h^2}{3} + ...
$$
This is the result of \cite{Dias:2011tj}. For $e^2>32/2$ all extremal RN black holes are superradiantly unstable.
\end{itemize}
It should be noted that the condition $e\mu >\Delta_0$ is obtained in a perturbative expansion in $r_h$ and, in consequence, it can't be extrapolated to sufficiently large RN black holes.

\section{Renormalized action}

The central quantity in the grand canonical ensemble is the grand potential $\Omega(T,\mu)$. The AdS/CFT correspondence identifies it with the on-shell regularized Euclidean action. To compute it we continue to Euclidean signature and  compactify time with a period $\frac{1}{T}$.
Writing $S_{on-shell} =\frac{\hat S_{bulk}}{T}$ one has
\be
\hat S_{bulk}=-\int d^3x \sqrt{-g}{\cal L}
\ee
where
\be
{\cal L}=\frac{1}{2\kappa^2}\left(R-2\Lambda\right)-D_\mu\phi \bar D^\mu\phi-\frac{1}{4} F_{\mu\nu} F^{\mu\nu}\, .
\ee
Using the equations of motion one can show that
$2{\cal L}=- (G_t^t+G_x^x)$ whence
\beqa
\hat S_{bulk}
&=&4\pi \left(\int_{x_h}^{x_b} dx \sec ^2(x) e^{-\delta(x)}-e^{-\delta(x)}\tan(x)\sec^2(x)f(x)\mid_{x=x_b}\right)\label{Seuc}
\eeqa
where the (diverging) result has been regularized at some boundary value $x_b = \pi/2-\epsilon$.

To have a well defined variational problem we must add the Gibbons-Hawking term
\beq
\hat S_{GH}=\int  d\theta d\phi \left.  \sqrt{\bar g} K \right\vert_{x=x_b}\, ,
\eeq
where $\bar g$ stands for the induced metric at $x_b$, $K=g^{\mu\nu}\nabla_{\mu\nu}n_\nu $
 is the extrinsic curvature   and $n^\mu  =\cos(x)\sqrt{f(x)}\delta^{\mu}{_x}$
 is the outward pointing unit normal vector to the boundary
 \be
\hat S_{GH}=-2\pi \left[\tan (x) e^{-\delta (x)} \left(\tan (x) f'(x)+f(x) \left(-2 \tan (x) \delta'(x)+6 \sec ^2(x)-2\right)\right)\right]_{x=x_b}\label{Sgh}\, .
\ee
The sum $\hat S_{bulk}+\hat S_{GH}$ diverges when $x_b\to \pi/2$. This can be handled by the following counterterm
\be
\hat S_{ct}=-4\int_{x=\pi/2-\epsilon} d\theta d\phi \sqrt{-g_{bdy}}=16\pi \left[\frac{\tan(x)^2}{\cos(x)}\sqrt{f(x)}e^{-\delta(x)}\right]_{x=x_b}\, .
\ee
 The grand canonical thermodynamic potential $\Omega$ is obtained from the limit $\epsilon\to 0$
\be
\Omega=\lim_{x\rightarrow\frac{\pi}{2}}\left(\hat S_{bulk}+\hat S_{GH}+\hat S_{ct}\right)  \, .\label{potential}
\ee

We checked the Smarr relations for this expression which, in particular,  imply that, indeed, $\Omega=M-T S-\mu Q$, as demanded by thermodynamic consistency. If we wish to compute the thermodynamic properties in the canonical ensemble (fixed $Q$) instead the relevant quantity to consider would be the free energy $F= M-T S$.

\section{Numerics}

\subsection{Time-evolution code}

\noindent  It is a virtue of the coordinate system \eqref{line1} that the Einstein and Maxwell equations appear as constraints \eqref{Einstein1}, \eqref{Einstein2} and \eqref{intat} that can be solved at each instant of time. The evolution of the system is then driven by the scalar field equations  \eqref{eqmovPhi} and \eqref{eqmovPi}. Starting from given nonequilibrium initial data, such as \eqref{gauss1} and \eqref{gauss2}, we have solved these equations numerically by resorting to a fourth-order accurate finite-difference evolution code. \\\\
\noindent Time evolution is performed by an explicit Runge-Kutta method. In order to deal with the high-frequency noise generated due to the finiteness of the discretization grid, we implement standard Kreiss-Oliger dissipation. By setting $\delta(\pi/2)=0$ we obtain stable evolutions with a constant Courant factor $\lambda$ with no need of local mesh refinement in time. On the other hand, spatial derivatives are discretized by employing a centered finite-difference stencil, while integrations are handled by a specifically designed routine, based on local polynomial interpolation. To deal with boundary conditions and numerical stability both at $x=0,\pi/2$ requires some detailed procedures that can be found in \cite{Maliborski:2013via}. \\\\
\noindent The major difficulty in the present setup stems from the fact that, upon evolution, the scalar profile develops very spiky features that demand a high resolution. To resolve these sharp features, which are apparent in figure \ref{fig:hairydecay}, we used global mesh refinement in space, eventually reaching $2^{17}+1$ grid points to discretize the interval  $x\in [0,\pi/2]$. This has required a parallel implementation that employs the MPI infrastructure to run the code on the SVG cluster at the CESGA facility (www.cesga.es). Optimal results have been obtained for $\sim 30$ nodes running in parallel. Smarter solutions involving local space mesh refinement are left for the future. \\\\
\noindent  The quality control parameters employed to activate the refinement process are both the norm of the momentum contraint \eqref{momentumconstraint}, as well as the relative mass loss at each time step. As a matter of fact, only at late times in the simulation are such mentioned fine resolutions required. The code stops at a time $t_f$ when the minimum value of $A(t,x)$ reaches below an user defined cutoff $A_c$.\footnote{For the simulations shown in the main text, we have set $A_c = 0.02$.} This is the time that is meant in the right figures \ref{fig:imecollapsesigma0d1e20} and \ref{fig:imecollapsesigma0d2e20}. Of course, mathematically speaking, the apparent horizon will only form in the infinite future $\lim_{t_f\to\infty} {\rm min} [A(t_f,x)]= 0$ in the chosen coordinate gauge.

\subsection{Convergence tests}

\subsubsection{Time evolution code}

\begin{figure}[h!]
\begin{center}
\includegraphics[width=16cm]{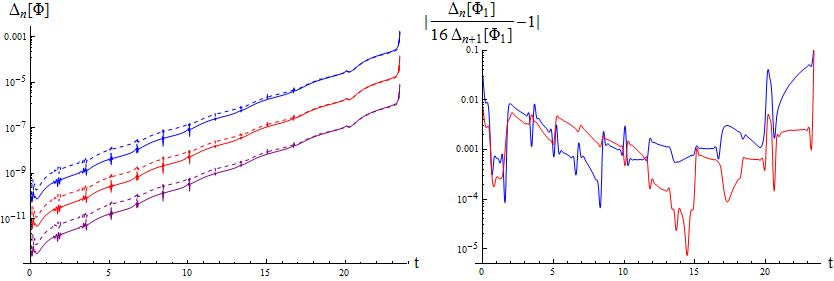}
\end{center}
\caption{\label{time_convergence} {\small Left: $\Delta_n[\Phi]$ for $n = 11$ (blue), $n = 12$ (red) and $n = 13$ (purple). Solid lines correspond to $\Phi_1$, while dashed ones to $\Phi_2$. The error norm decreases with increasing grid resolution. Right:  absolute value deviation of the quotient $\frac{\Delta_n[\Phi_1]}{2^4 \Delta_{n+1}[\Phi_1]}$ from 1 for $n = 11$ (blue) and $n = 12$ (red). The error norm converges to zero at the right order.}}
\end{figure}

\noindent To illustrate the convergence properties of our time evolution code, we consider the functional
\beq
\Delta_n[g](t) \equiv \| g_n(t,x) - g_{n+1}(t,x) \| = \left( \int_0^{\pi/2} \tan(x)^2 (g_n(t,x) - g_{n+1}(t,x) )^2 \right)^{1/2},
\eeq
where $g_n$ refers to any function computed on a discretization grid of spatial resolution $h = \pi/2^{n+1}$. Fourth-order convergence then implies that
\beq
\Delta_{n+1}[g](t) = 2^{-4} \Delta_n[g](t). \label{4th_order_test}
\eeq
In figure \ref{time_convergence}, we plot $\Delta_n[\Phi_1](t), \Delta_n[\Phi_2](t)$ at $n = 11, 12, 13$ for simulations with initial data
\beqa
\Phi &=& \epsilon_1 \frac{2}{\pi} \sin x \exp\left( -\frac{4 \tan^2(x)}{\pi^2 \sigma^2} \right), \\
\Pi &=& i \epsilon_2 \frac{2}{\pi} \exp \left( -\frac{4\tan^2(x)}{\pi^2 \sigma^2}  \right),
\eeqa
where $\epsilon_1 = \epsilon_2 = 12$, $\sigma = 0.1$ and $e = 5$. The scalar field completes seven bounces before collapse is achieved at $t_f = 23.47$. We see precisely that relation \eqref{4th_order_test} is fulfield.\footnote{We have deactivated the global mesh refinement algorithm, and thus we observe deviations from exact fourth-order convergence right before collapse.}

\subsubsection{Soliton eigenfrequencies pseudospectral code}
\begin{figure}[h!]
\begin{center}
\includegraphics[width=16cm]{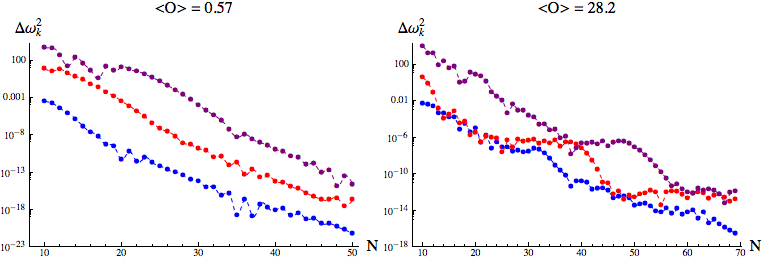}
\end{center}
\caption{\label{convergence_soliton_normal_modes} {\small Left: At $e = 5$, and for a soliton with $\left<\mathcal{O}\right> = 0.57$, we plot the errors $\Delta\omega_1^2$ (blue), $\Delta\omega_{10}^2$ (red), $\Delta\omega_{20}^2$ (purple) on a collocation grid ranging from $N = 10$ to $N = 50$. It is seen that the error tends exponentially to zero as the resolution $N$ is increased. Right: The same quantities for the e = 5, $\left<\mathcal{O}\right> = 28.2$ soliton, on a collocation grid ranging from $N = 10$ to $N = 70$.}}
\end{figure}

\noindent For the computation of the normal modes  on top of a soliton background, we have resorted to the pseudospectral method described  in the main text. As mentioned, the output of this procedure are the first $N$ soliton normal modes.  For pseudospectral methods, we expect exponential convergence, since we are approximating the analytical scalar field eigenmodes. In order to determine the convergence properties of the method, we show in figure \ref{convergence_soliton_normal_modes} the quantity $\Delta\omega_k^2(N)$, defined as
\beq
\Delta\omega_k^2(N) = \left| \omega_k^2(N+1) - \omega_k^2(N) \right|.
\eeq
The fact that $\Delta\omega_k^2(N) \rightarrow 0$ exponentially as $N \rightarrow \infty$  implies that the sequence $\{\omega_k^2(N), N=N_0, N_0+1,...\}$ converges as anticipated.
\\\\

%%%%%%%%%%%%%%%%%%%%%%%%%%%%%%%%%%%%%%%%%%%%%%%%%%%%%%%%%%%%%%%%%%%%%%%%%%

\end{document}